\documentclass[11pt]{article} 
\usepackage{amsmath}
\usepackage{amssymb} 
\usepackage{amsthm} 
\usepackage{amsfonts}
\usepackage{fancyheadings} 
\usepackage{graphics} 
\usepackage{psfrag}
\usepackage{graphicx,subfigure}

\newtheorem{theorem}{Theorem}
\newtheorem{lemma}[theorem]{Lemma}
\newtheorem{proposition}[theorem]{Proposition}

\newtheorem{definition}[theorem]{Definition}

\makeatletter \long\def\@makecaption#1#2{
        \vskip 0.8ex
        \setbox\@tempboxa\hbox{\small {\bf #1:} #2}
        \parindent 1.5em  
        \dimen0=\hsize
        \advance\dimen0 by -3em
        \ifdim \wd\@tempboxa >\dimen0
                \hbox to \hsize{
                        \parindent 0em
                        \hfil
                        \parbox{\dimen0}{\def\baselinestretch{0.96}\small
                                {\bf #1.} #2
                                }                         \hfil}
        \else \hbox to \hsize{\hfil \box\@tempboxa \hfil}
        \fi
        }
\makeatother

\newcommand{\vtiny}{\vspace*{.05in}}
\long\def\comment#1{}

\makeatletter
\def\@cite#1#2{[\if@tempswa #2 \fi #1]}
\makeatother

\newcommand{\widgraph}[2]{\includegraphics[keepaspectratio,width=#1]{#2}}

\newcommand{\xvec}{\ensuremath{x}}
\newcommand{\xvar}{\ensuremath{x}}
\newcommand{\sset}{\ensuremath{S}}
\newcommand{\defn}{\ensuremath{:  =}}
\newcommand{\clauseset}{\ensuremath{C}}
\newcommand{\cnum}{\ensuremath{m}}
\newcommand{\vertexset}{\ensuremath{V}}
\newcommand{\vnum}{\ensuremath{n}}

\newcommand{\edge}{\ensuremath{E}}

\newcommand{\acl}{\ensuremath{a}}
\newcommand{\bcl}{\ensuremath{b}}
\newcommand{\ccl}{\ensuremath{c}}
\newcommand{\dcl}{\ensuremath{d}}

\newcommand{\E}{{\rm E}}

\newcommand{\graph}{\ensuremath{G}}

\newcommand{\Nbrcl}{\ensuremath{\vertexset}}

\newcommand{\Nbrvar}{\ensuremath{\clauseset}}

\newcommand{\Nbrvarpos}{\ensuremath{\Nbrvar^+}}
\newcommand{\Nbrvarneg}{\ensuremath{\Nbrvar^-}}

\newcommand{\Nbrvarzero}{\ensuremath{\Nbrvarneg}}
\newcommand{\Nbrvarone}{\ensuremath{\Nbrvarpos}}

\newcommand{\Nbragree}[2]{\ensuremath{\Nbrvar_{#1}^s(#2)}}
\newcommand{\Nbrdis}[2]{\ensuremath{\Nbrvar_{#1}^u(#2)}}

\newcommand{\Jsca}{\ensuremath{J}}

\newcommand{\compat}{\ensuremath{\psi}}
\newcommand{\clfun}[1]{\ensuremath{\compat_{\Jsca_#1}}}

\newcommand{\Messtovar}{\ensuremath{\eta}}
\newcommand{\messtovar}[2]{\ensuremath{\Messtovar_{#1 \ra #2}}}

\newcommand{\Messtocl}{\ensuremath{\Pi}}
\newcommand{\messtocl}[2]{\ensuremath{\Messtocl_{#1 \ra #2}}}
\newcommand{\messtoclun}[2]{\ensuremath{\Messtocl^u_{#1 \ra #2}}}
\newcommand{\messtoclsat}[2]{\ensuremath{\Messtocl^s_{#1 \ra #2}}}
\newcommand{\messtoclstar}[2]{\ensuremath{\Messtocl^*_{#1 \ra #2}}}

\newcommand{\Bias}{\ensuremath{B}}

\newcommand{\spmarg}{\ensuremath{\mu}}

\newcommand{\biasfrac}{\ensuremath{\beta}}

\newcommand{\sat}[2]{\ensuremath{s_{#1,#2}}}
\newcommand{\unsat}[2]{\ensuremath{u_{#1,#2}}}

\newcommand{\unc}{\ensuremath{o}}

\newcommand{\Setstar}{\ensuremath{S_*}}
\newcommand{\Setcon}{\ensuremath{S_c}}
\newcommand{\Setunc}{\ensuremath{S_\unc}}

\newcommand{\weight}{\ensuremath{\omega}}
\newcommand{\wun}{\ensuremath{\weight_\unc}}
\newcommand{\wcon}{\ensuremath{\weight_c}}
\newcommand{\wstar}{\ensuremath{\weight_*}}
\newcommand{\Weight}{\ensuremath{W}}

\newcommand{\Par}[1]{\ensuremath{P_{#1}}}
\newcommand{\Suppar}{\ensuremath{\mathcal{P}}}

\newcommand{\extX}{\ensuremath{\mathcal{X}}}

\newcommand{\supcompat}{\ensuremath{\Psi}}

\newcommand{\Val}[2]{\ensuremath{\operatorname{VAL}_{#1}(#2)}}

\newcommand{\Con}[2]{\ensuremath{\operatorname{CON}_{#1,#2}}}

\newcommand{\superp}{\ensuremath{p_{gen}}}

\newcommand{\Genmess}{\ensuremath{M}}
\newcommand{\tovarun}[2]{\ensuremath{\Genmess_{#1 \ra #2}^{u}}}
\newcommand{\tovarsat}[2]{\ensuremath{\Genmess_{#1 \ra #2}^{s}}}
\newcommand{\tovarstar}[2]{\ensuremath{\Genmess_{#1 \ra #2}^{*}}}

\newcommand{\Toclmess}{\ensuremath{M}}

\newcommand{\Rsat}[2]{\ensuremath{R_{#1 \ra #2}^s}}
\newcommand{\Run}[2]{\ensuremath{R_{#1 \ra #2}^u}}
\newcommand{\Rstar}[2]{\ensuremath{R_{#1 \ra #2}^*}}

\newcommand{\SP}{\ensuremath{\operatorname{SP}}}

\newcommand{\MRF}{\ensuremath{\operatorname{MRF}}}

\newcommand{\eps}{\epsilon}

\newcommand{\Field}{\ensuremath{F}}

\newcommand{\Core}[2]{\ensuremath{\gamma_{#1}(#2)}}

\newcommand{\Consistent}[1]{\ensuremath{A}}

\newcommand{\Reachable}[1]{\ensuremath{B}}

\newcommand{\MaptoReachable}{\sigma}
\newcommand{\MapfromReachable}{\tau}

\newcommand{\Ind}[1] {{\rm Ind} [#1]}

\newcommand{\ra}{\rightarrow}

\newcommand{\dirgraph}{\ensuremath{G}}

\newcommand{\Kform}{\ensuremath{\phi}}
\newcommand{\csize}{\ensuremath{C}}
\newcommand{\crat}{\ensuremath{c}}

\newcommand{\Prob}{\ensuremath{\mathbb{P}}}

\newcommand{\Mezard}{M\'{e}zard}

\usepackage{ifthen}

\begin{document}

\newcommand{\doctype}{TECH}

\ifthenelse{\equal{\doctype}{TECH}} 
{\typeout{----------> Technical report formatting}}
{\typeout{----------> SODA formatting}}

\ifthenelse{\equal{\doctype}{TECH}} 
{ \bibliographystyle{abbrv}}
{ \bibliographystyle{latex8}}


\newcommand{\fn}{\footnotesize}
\renewcommand{\thefootnote}{\arabic{footnote}}

{\bf} \title{A New Look at Survey Propagation and its Generalizations}

\author{Elitza Maneva\footnote{Department of Electrical Engineering
and Computer Science, UC Berkeley, CA.  Email:
{\texttt{elitza@eecs.berkeley.edu}}.  Supported by NSF grant
CCR-0121555.}  \and
Elchanan Mossel\footnote{Department of Statistics, UC Berkeley, CA.
Email: {\texttt{mossel@stat.berkeley.edu}}.  Supported by a Miller
Fellowship in Computer Science and Statistics, NSF grant DMS-0504245 and a Sloan Fellowship in Mathematics}  \and
Martin J. Wainwright\footnote{Department of Electrical Engineering and
Computer Science and Department of Statistics, UC Berkeley, CA.
Email: {\texttt{wainwrig@eecs.berkeley.edu}}. Supported by a Sloan
Fellowship in Computer Science and a grant from Intel Corporation.  }
}

\maketitle

\renewcommand{\thefootnote}{\arabic{footnote}}

\begin{center}
\begin{abstract}
This paper provides a new conceptual perspective on \emph{survey
propagation}, which is an iterative algorithm recently introduced by
the statistical physics community that is very effective in solving
random $k$-SAT problems even with densities close to the
satisfiability threshold.  We first describe how any SAT formula can
be associated with a novel family of Markov random fields (MRFs),
parameterized by a real number $\rho \in [0,1]$.  We then show that
applying belief propagation---a well-known ``message-passing''
technique for estimating marginal probabilities---to this family of
MRFs recovers a known family of algorithms, ranging from pure survey
propagation at one extreme ($\rho = 1$) to standard belief propagation
on the uniform distribution over SAT assignments at the other extreme
($\rho = 0$).  Configurations in these MRFs have a natural
interpretation as partial satisfiability assignments, on which a
partial order can be defined.  We isolate \emph{cores} as minimal
elements in this partial ordering, which are also fixed points of
survey propagation and the only assignments with positive probability
in the MRF for $\rho=1$.  Our experimental results for $k=3$ suggest
that solutions of random formulas typically do not possess non-trivial
cores.  This makes it necessary to study the structure of the space of
partial assignments for $\rho<1$ and investigate the role of
assignments that are very close to being cores.  To that end, we
investigate the associated lattice structure, and prove a
weight-preserving identity that shows how any MRF with $\rho > 0$ can
be viewed as a ``smoothed'' version of the uniform distribution over
satisfying assignments ($\rho=0$).  Finally, we isolate properties of
Gibbs sampling and message-passing algorithms that are typical for an
ensemble of $k$-SAT problems.

\vspace*{.2in}


\end{abstract}
\end{center}

\newpage

\section{Introduction}

Constraint satisfaction problems play an important role across a broad
spectrum of computer science, including computational complexity
theory~\cite{Cook71}, coding theory~\cite{Gallager63,Richardson01a},
and artificial intelligence~\cite{Pearl88,Dechter03}.  Important but
challenging problems include devising efficient algorithms for finding
satisfying assignments (when the problem is indeed satisfiable), or
conversely providing a certificate of unsatisfiability.  One of the
best-known examples of a constraint satisfaction problem is the
$k$-SAT problem, which is a classical NP complete
problem~\cite{Cook71} for all $k \geq 3$.  In trying to understand the
origin of its hardness, a great deal of research has been devoted to
the properties of formulas drawn from different probability
distributions. One of the most natural models for random $k$-SAT
problems is the following: for a fixed density parameter $\alpha >0$,
choose $\cnum = \alpha \vnum$ clauses uniformly and with replacement
from the set of all $k$-clauses on $\vnum$ variables.  Despite its
simplicity, many essential properties of this model are yet to be
understood: in particular, the hardness of deciding if a random
formula is satisfiable and finding a satisfying assignment for a
random formula are both major open
problems~\cite{Levin86,Wang97,Feige02}.

One of the most exciting recent developments in satisfiability
problems has its origins not in computer science, but rather in
statistical physics.  More specifically, the ground-breaking
contribution of \Mezard, Parisi and Zecchina~\cite{MPZ02}, as
described in an article published in ``Science'', is the development
of a new algorithm for solving $k$-SAT problems.  A particularly
dramatic feature of this method, known as \emph{survey propagation},
is that it appears to remain effective at solving very large instances
of random $k$-SAT problems----even with densities very close to the
satisfiability threshold, a regime where other ``local'' algorithms
(e.g., the WSAT method~\cite{SKC93}) typically fail.  Given this
remarkable behavior, the survey propagation algorithm has generated a
great deal of excitement and follow-up work in both the statistical
physics and computer science
communities~\cite[e.g.,]{BMZ03,BMWZ03,BZ04,Aurell05,AR05,MMZ05,Parisi02,WM05}.
Nonetheless, despite the considerable progress to date, the reasons
underlying the remarkable performance of survey propagation are not
yet fully understood.

\subsection{Our contributions}

This paper provides a novel conceptual perspective on survey
propagation---one that not only sheds light on the reasons underlying
its success, but also places it within a broader framework of related
``message-passing'' algorithms that are widely used in different
branches of computer science.  More precisely, by introducing a new
family of Markov random fields (MRFs) that are associated with any
$k$-SAT problem, we show how a range of algorithms---including survey
propagation as a special case---can all be recovered as instances of
the well-known belief propagation algorithm~\cite{Pearl88}, as applied
to suitably restricted MRFs within this family.  This equivalence is
important because belief propagation is a message-passing
algorithm---widely used and studied in various areas, including coding
theory~\cite{Richardson01a,Frank01,Yedidia05}, computer
vision~\cite{Freeman00,Coughlan02} and artificial
intelligence~\cite{Pearl88,YFW03}----for computing approximations to
marginal distributions in Markov random fields.  Moreover, this
equivalence motivates a deeper study of the combinatorial properties
of the family of extended MRFs associated with survey propagation.
Indeed, one of the main contributions of our work is to reveal the
combinatorial structures underlying the survey propagation algorithm.

\begin{figure}
\begin{center}
\widgraph{.5\textwidth}{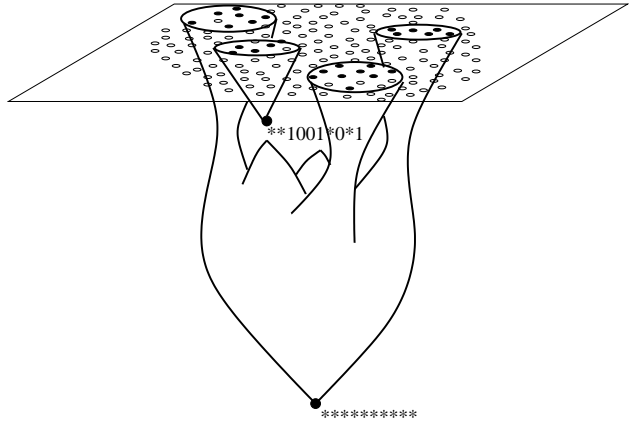} 
\caption{The set of fully assigned satisfying configurations occupy
the top plane, and are arranged into clusters.  Enlarging to the space
of partial assignments leads to a new space with better connectiviity.
Minimal elements in the partial ordering are known as cores. Each
core corresponds to one or more clusters of solutions from the top
plane. In this example, one of the clusters has as a core a
non-trivial partial assignment, whereas the others are connected to
the all-$\ast$ assignment.}
\label{fig:space}
\end{center}
\end{figure}

The configurations in our extended MRFs turn out to have a natural
interpretation as particular types of \emph{partial SAT assignments},
in which a subset of variables are assigned $0$ or $1$ variables in
such a way that the remaining formula does not contain any empty or
unit clauses.  To provide some geometrical intuition for our results,
it is convenient to picture these partial assignments as arranged in
layers depending on the number of assigned variables, so that the top
layer consists of fully assigned satisfying configurations.
Figure~\ref{fig:space} provides an idealized illustration of the space
of partial assignments viewed in this manner.  It is
argued~\cite{MRZ03,MMZ05,AR05} that for random formulas with high
density of clauses, the set of fully assigned configurations are
separated into disjoint clusters that cause local message-passing
algorithms like belief propagation to break down (see
Figure~\ref{fig:clustering} for an illustration). Based on our
results, the introduction of partial SAT assignments yields a
\emph{modified search space} that is far less fragmented, thereby
permitting a local algorithm like belief propagation to find
solutions.

We show that there is a natural partial ordering associated with this
enlarged space, and we refer to minimal elements in this partial
ordering as \emph{cores}.  We prove that any core is a fixed point of
the pure form of survey propagation ($\rho = 1$). This fact indicates
that each core represents a summary of one cluster of
solutions. However, our experimental results for $k=3$ indicate that
the solutions of random formulas typically have trivial cores (i.e.,
the empty assignment).  This observation motivates deeper study of the
full family of Markov random fields for the range $0 \leq \rho \leq
1$, as well as the associated belief propagation algorithms, which we
denote by $\SP(\rho)$.  Accordingly, we study the lattice structure of
the partial assignments, and prove a combinatorial identity that
reveals how the distribution for $\rho \in (0,1)$ can be viewed as a
``smoothed'' version of the MRF with $\rho = 0$.  Our experimental
results on the $\SP(\rho)$ algorithms indicate that they are most
effective for values of $\rho$ close to and not necessarily equal to
$1$. One intriguing possibility is that the effectiveness of pure
survey propagation (i.e., $\SP(1)$) may be a by-product of the fact
that $\SP(\rho)$ is most effective for values of $\rho$ less than $1$,
but going to $1$ as $n$ goes to infinity. The near-core assignments
which are the ones of maximum weight in this case, may correspond to
quasi-solutions of the cavity equations, as defined by
Parisi~\cite{Parisi02}.
In addition, we consider alternative sampling-based methods (e.g.,
Gibbs sampling) for computing marginals for the extended MRFs.  We
also study properties of both message-passing and Gibbs sampling that
are typical over a random ensemble of $k$-SAT problems.  We establish
results that link the typical behavior of Gibbs sampling and
message-passing algorithms under suitable initialization, and when
applied to the extended family of MRFs with $\rho$ sufficiently close
to one.

The fact that the pure form of survey propagation (i.e., $\SP(1)$ in
our notation) is a form of belief propagation was first conjectured by
Braunstein et al.~\cite{BMZ03}, and established independently of our
work by Braunstein and Zecchina~\cite{BZ04}.  In other independent
work, Aurell et al.~\cite{Aurell05} provided an alternative derivation
of $\SP(1)$ that established a link to belief propagation. However,
both of these papers treat only the case $\rho = 1$, and do not
provide a combinatorial interpretation based on an underlying Markov
random field.  The results established here are a strict
generalization, applying to the full range of $\rho \in [0,1]$.
Moreover, the structures intrinsic to our Markov random
fields---namely cores and lattices---highlight the importance of
values $\rho \neq 1$, and place the survey propagation algorithm on a
combinatorial ground.  As we discuss later, this combinatorial
perspective has already inspired subsequent work~\cite{AR05} on survey
propagation for satisfiability problems. Looking forward, the
methodology of partial assignments may also open the door to other
problems where a complicated landscape prevents local search
algorithms from finding good solutions.  As a concrete example, a
subset of the current authors~\cite{WM05} have recently shown that
related ideas can be leveraged to perform lossy data compression at
near-optimal (Shannon limit) rates.

\subsection{Organization}

The remainder of this paper is organized in the following way:

\begin{itemize}
\item In Section~\ref{subsec:lit}, we provide further background on
the $k$-SAT problem, as well as previous work on survey propagation.

\item In Section~\ref{sec:back}, we introduce required notation and
set up the problem more precisely.

\item In Section~\ref{sec:MRF}, we define a family of Markov random
fields (MRFs) over partial satisfiability assignments, and prove that
survey propagation and related algorithms correspond to belief
propagation on these MRFs.  

\item Section~\ref{SecCore} is devoted to analysis of the
combinatorial properties of this family of extended MRFs, as well as
some experimental results on cores and Gibbs sampling.
\item
In Section~\ref{SecRandEns}, we consider properties of random
ensembles of SAT formulae, and prove results that link the performance
of survey propagation and Gibbs sampling to the choice of Markov
random field.
\item We conclude with a discussion in Section~\ref{SecConclusion}.
\end{itemize}

\noindent We note that many of the results reported here have been
presented (without proofs or details) as an extended SODA
abstract~\cite{MMW05}.

\subsection{Previous work on $k$-SAT and survey propagation} 
\label{subsec:lit}
As a classical NP complete problem~\cite{Cook71}, the $k$-SAT problem
for $k \geq 3$ has been extensively studied.  One approach is to
consider ensembles of random formulas; in particular, a commonly
studied ensemble is based on choosing $\cnum = \alpha \vnum$ clauses
uniformly and with replacement from the set of all
$k$-clauses on $\vnum$ variables.  Clearly, a formula drawn randomly
from this ensemble becomes increasingly difficult to satisfy as the
clause density $\alpha > 0$ increases.  There is a large body of
work~\cite{Friedgut99,Goe96,CR92,Fer92,DBM00,KKL00,AP03} devoted to
the study of the {\em threshold} density where the formula becomes
unsatisfiable; however, except for the case $k=2$, the value of the
threshold is currently unknown.  However, non-rigorous techniques from
statistical physics can be applied to yield estimates of the
threshold; for instance, results from \Mezard$\,$ and
Zecchina~\cite{MZ97} yield a threshold estimate of $\alpha_c \approx
4.267$ for $k=3$.

The survey propagation (SP) algorithm, as introduced by \Mezard,
Parisi and Zecchina~\cite{MPZ02}, is an iterative message-passing
technique that is able to find satisfying assignments for large
instances of SAT problems at much higher densities than previous
methods.  The derivation of SP is based on the cavity method in
conjunction with the 1-step replica summetry breaking (1-RSB) ansatz
of statistical physics.  We do not go into these ideas in depth here,
but refer the reader to the physics literature~\cite{MZ02,BMZ03,MPZ02}
for further details.  In brief, the main assumption is the existence
of a critical value $\alpha_d$ for the density, smaller than the
threshold density $\alpha_c$, at which the structure of the space of
solutions of a random formula changes. For densities below $\alpha_d$
the space of solutions is highly connected---in particular, it is
possible to move from one solution to any other by single variable
flips,\footnote{There is no general agreement on whether assignments
should be considered neighbors if they differ in only one variable, or
any constant number of variables} staying at all times in a satisfying
assignment. 
\begin{figure}
\begin{center}
\begin{tabular}{ccccc}
\widgraph{.25\textwidth}{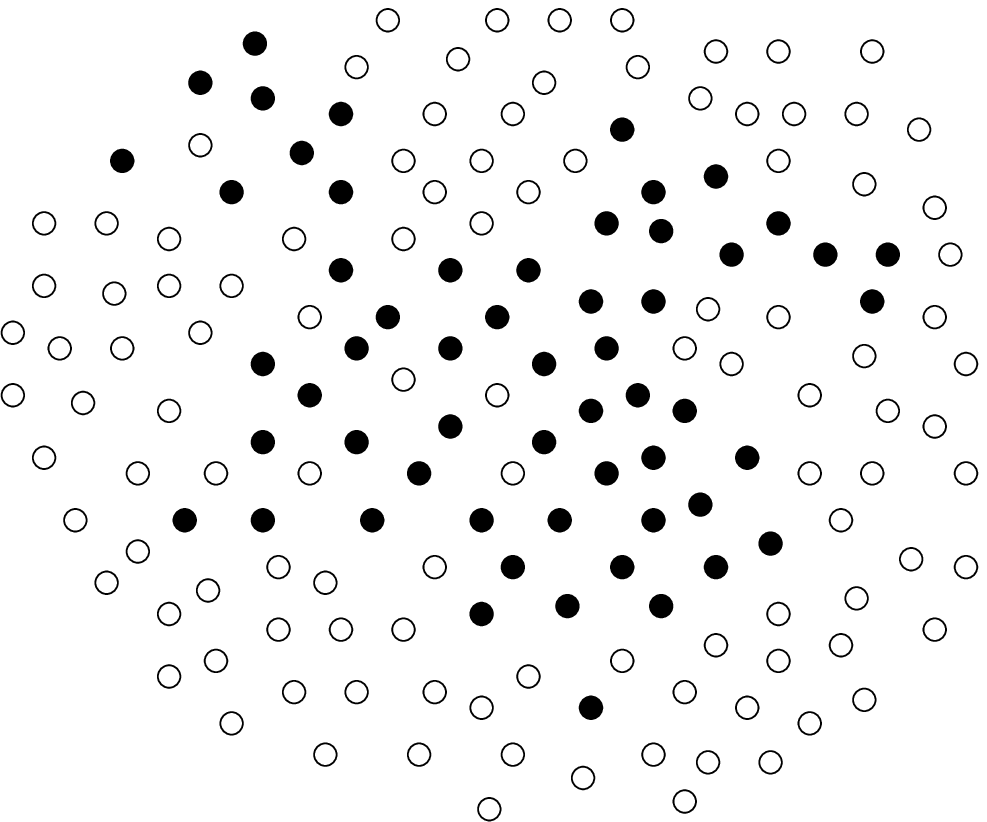} &
\hspace*{.1in} &
\widgraph{.3\textwidth}{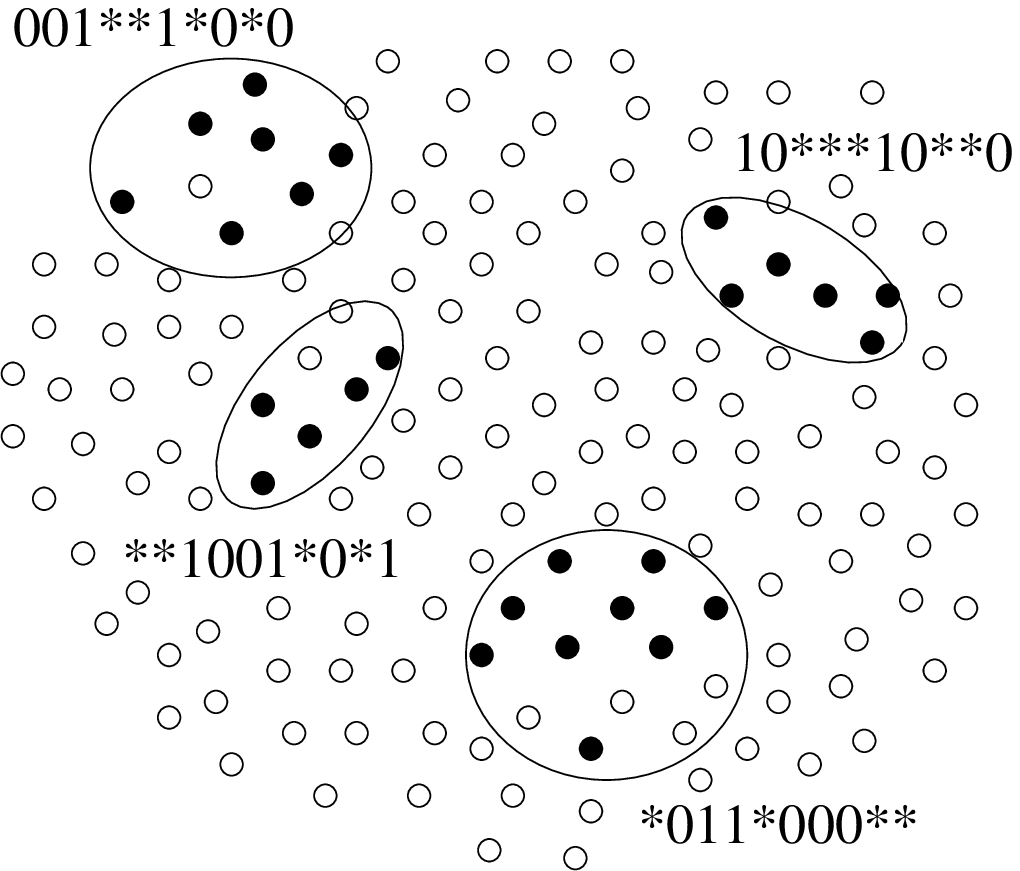} &
&
\widgraph{.25\textwidth}{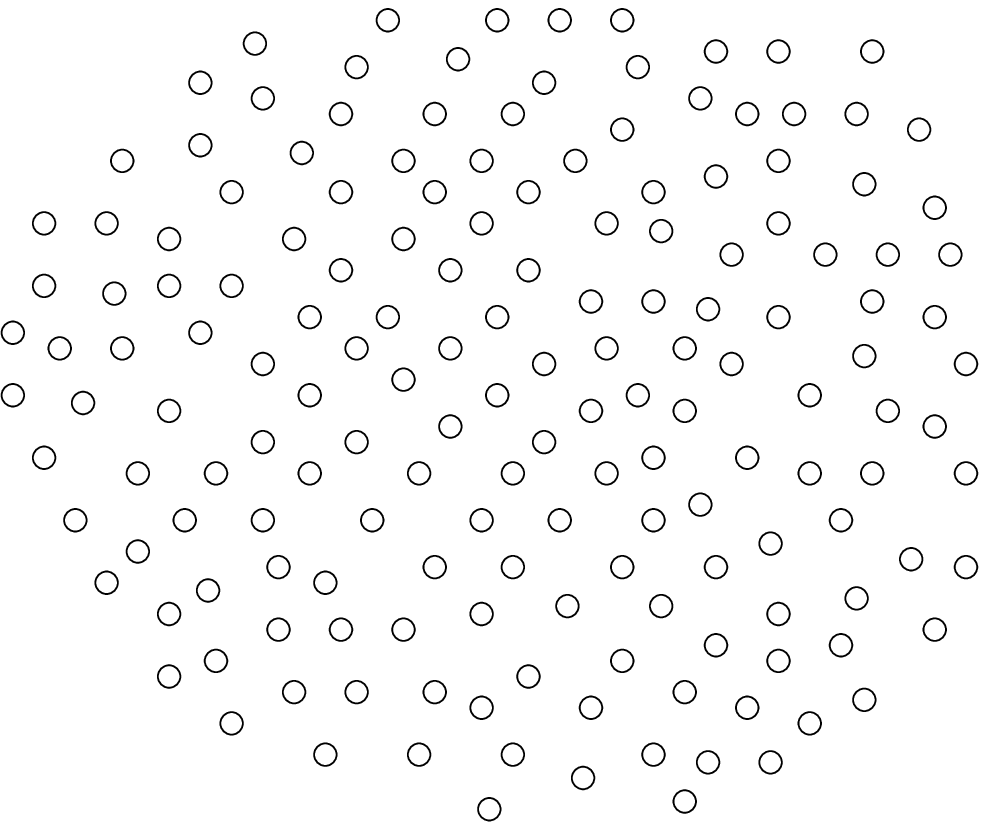} \\
(a) $0<\alpha<\alpha_d$ & &
(b) $\alpha_d<\alpha<\alpha_c$& & 
(c) $\alpha_c<\alpha$ \\
&&&&\\
\end{tabular}
\caption{The black dots represent satisfying assignments, and white
dots unsatisfying assignments. Distance is to be interpreted as the
Hamming distance between assignments. (a) For low densities the space
of satisfying assignments is well connected. (b) As the density
increases above $\alpha_d$ the space is believed to break up into an
exponential number of clusters, each containing an exponential number
of assignments. These clusters are separated by a ``sea'' of
unsatisfying assignments. (c) Above $\alpha_c$ all assignments become
unsatisfying. }
\label{fig:clustering}
\end{center}
\end{figure}
 For densities above $\alpha_d$, the space of solutions
breaks up into clusters, so that moving from a SAT assignment within
one cluster to some other assignment within another cluster requires
flipping some constant fraction of the variables
simultaneously. Figure~\ref{fig:clustering} illustrates how the
structure of the space of solutions evolves as the density of a random
formula increases.  The clustering phenomenon that is believed to
occur in the second phase is known in the statistical physics
literature as 1-step replica symmetry breaking~\cite{MZ02}, and the
estimated value for $\alpha_d$ in the case $k=3$ is $\alpha_d \approx
3.921$.  Within each cluster, a distinction can be made between
\emph{frozen} variables---ones that do not change their value within
the cluster---and \emph{free} variables that do change their value in
the cluster. A concise description of a cluster is an assignment of
$\{0, 1, *\}$ to the variables with the frozen variables taking their
frozen value, and the free variables taking the joker or wild card
value $*$.  The original argument for the clustering assumption was
the analysis of simpler satisfiability problems, such as XOR-SAT,
where the existence of clusters can be demonstrated by rigorous
methods~\cite{MRZ03}. In addition, if one assumes that there are no
clusters, the cavity method calculation yields a value for
$\alpha_c>5$ (for $k=3$), which is known to be wrong.  More recently,
Mora, M\'{e}zard and Zecchina~\cite{MMZ05} have demonstrated via
rigorous methods that for $k\ge 8$ and some clause density below the
unsatisfiability threshold, clusters of solutions do indeed exist.

The survey propagation (SP) algorithm is so-named, because like the
belief propagation algorithm~\cite{Pearl88, YFW03}, it entails
propagating statistical information in the form of messages between
nodes in the graph.  In the original derivation of the
updates~\cite{MPZ02,BMZ03}, the messages are interpreted as
``surveys'' taken over the clusters in solution space, which provide
information about the fraction of clusters in which a given variable
is free or frozen.  However, prior to the work presented here, it was
not clear how to interpret the algorithm as an instantiation of belief
propagation, and thus as a method for computing (approximations) to
marginal distributions in a certain Markov random field (MRF).
Moreover, as discussed above, our formulation of SP in this manner
provides a broader view, in which SP is one of many possible
message-passing algorithms that can be applied to smoothed MRF
representations of SAT problems.

\section{Background and problem set-up}
\label{sec:back}

In this section, we begin with notation and terminology necessary to
describe the $k$-SAT problem, and then provide a precise description
of the survey propagation updates.

\subsection{The $k$-SAT problem and factor graphs}
\label{subsec:sat}

\paragraph{Basic notation:} Let $\clauseset$ and $\vertexset$ 
represent index sets for the clauses and variables, respectively,
where $|\vertexset| = \vnum$ and $|\clauseset| = \cnum$.  We denote
elements of $\vertexset$ using the letters $i,j,k$, etc., and members
of $\clauseset$ with the letters $\acl, \bcl, \ccl$, etc.  We use
$\xvec_\sset$ to denote the subset of variables $\{\xvar_i: \; i \in
\sset\}$.  In the $k$-SAT problem, the clause indexed by $\acl \in
\clauseset$ is specified by the pair $(\Nbrcl(\acl), \Jsca_\acl)$,
where $\Nbrcl(\acl) \subset \vertexset$ consists of $k$ elements, and
$\Jsca_\acl \defn ( \Jsca_{\acl, i}: \; i \in \Nbrcl(\acl) )$ is a
$k$-tuple of $\{0,1\}$-valued weights.  The clause indexed by $\acl$
is \emph{satisfied} by the assignment $\xvec$ if and only if
$\xvec_{\Nbrcl(\acl)} \neq \Jsca_\acl$.
Equivalently, letting $\delta(y,z)$ denote an indicator function for
the event $\{y=z\}$, if we define the function
\begin{eqnarray}
\label{EqnDefnCompat}
\clfun{\acl}(\xvec) & \defn & 1- \prod_{i \in \Nbrcl(\acl)}
\delta(\Jsca_{\acl, i}, \xvar_i),
\end{eqnarray}
then the clause $\acl$ is satisfied by $\xvec$ if and only if
$\clfun{\acl}(\xvec) = 1$.  The overall formula consists of the AND of
all the individual clauses, and is satisfied by $\xvec$ if and only if
$\prod_{\acl \in \clauseset} \clfun{\acl}(\xvec) = 1$.

\begin{figure}[h]
\begin{center}
\psfrag{#a#}{\acl} \psfrag{#b#}{\bcl} \psfrag{#c#}{\ccl}
\psfrag{#d#}{\dcl} \psfrag{#1#}{$1$} \psfrag{#2#}{$2$}
\psfrag{#3#}{$3$} \psfrag{#4#}{$4$} \psfrag{#5#}{$5$}
\widgraph{.25\textwidth}{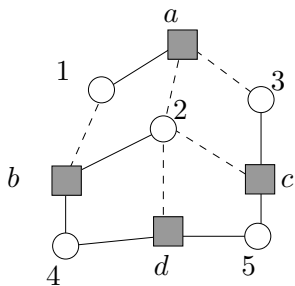}
\caption{Factor graph representation of a $3$-SAT problem on $\vnum =
5$ variables with $\cnum = 4$ clauses, in which circular and square
nodes correspond to variables and clauses respectively.  Solid and
dotted edges $(\acl, i)$ correspond to the weightings $\Jsca_{\acl,i}
= 0$ and $\Jsca_{\acl, i} = 1$ respectively.  The clause $\acl$ is
defined by the neighborhood set $\Nbrcl(\acl) = \{1,2,3\}$ and weights
$\Jsca_{a} = (0,1,1)$. In traditional notation, this corresponds to
the formula $(x_1 \vee \bar{x}_2 \vee \bar{x}_3) \wedge (\bar{x}_1
\vee x_2 \vee x_4) \wedge (\bar{x}_2 \vee x_3 \vee x_5) \wedge
(\bar{x}_2 \vee x_4 \vee x_5)$.}
\label{FigKsat}
\end{center}
\end{figure}

\paragraph{Factor graphs:} A convenient graphical representation of 
any $k$-SAT problem is provided by the formalism of factor graphs
(see~\cite{Frank01} for further background).  As illustrated in
Figure~\ref{FigKsat}, any instance of the $k$-SAT problem can be
associated with a particular bipartite graph on the variables (denoted
by circular nodes) and clauses (denoted by square nodes), where the
edge $(\acl, i)$ between the clause $\acl \in \clauseset$ and variable
$i \in \vertexset$ is included in $\edge$ if and only if $i \in
\Nbrcl(\acl)$.  Following Braunstein et al.~\cite{BMZ03}, it is
convenient to introduce two labellings of any given edge---namely,
solid or dotted, corresponding to whether $\Jsca_{\acl,i}$ is equal to
$0$ or $1$ respectively.

For later use, we define (for each $i \in \vertexset$) the set
$\Nbrvar(i) \defn \{\acl \in \clauseset \, : \, i \in \Nbrcl(\acl)\}$,
corresponding to those clauses that impose constraints on variable
$x_i$.  This set of clauses can be decomposed into two disjoint
subsets
\begin{equation}
\begin{array}{ll}
\label{EqnDefnNbrvar}
\Nbrvarneg(i) \defn \{\acl \in \Nbrvar(i) \; : \; \Jsca_{\acl,
i}=1\},
 &
\qquad \Nbrvarpos(i) \defn \{\acl \in \Nbrvar(i) \; : \; \Jsca_{\acl,
i}=0\},
\end{array}
\end{equation}
according to whether the clause is satisfied by $x_i = 0$ or $x_i = 1$
respectively.  Moreover, for each pair $(\acl, i) \in \edge$, the set
$\Nbrvar(i) \backslash \{\acl\}$ can be divided into two (disjoint)
subsets, depending on whether their preferred assignment of $\xvar_i$
\emph{agrees} (in which case $\bcl \in\Nbragree{\acl}{i}$) or
\emph{disagrees} (in which case $\bcl \in \Nbrdis{\acl}{i}$) with the
preferred assignment of $\xvar_i$ corresponding to clause $\acl$.
More formally, we define
\begin{equation}
\label{EqnDefnNbragreedis}
\begin{array}{ll}
\Nbragree{\acl}{i} \defn \{ \bcl \in \Nbrvar(i) \backslash \{\acl\} \;
: \; \Jsca_{\acl, i} = \Jsca_{\bcl, i} \; \},&
\qquad \Nbrdis{\acl}{i} \defn \{ \bcl \in \Nbrvar(i) \backslash
\{\acl\} \; : \; \Jsca_{\acl, i} \neq \Jsca_{\bcl, i} \; \}.
\end{array}
\end{equation}

Our focus is on random ensembles of $k$-SAT instances: for a given
clause density $\alpha >0$, a random instance is obtained by sampling
$\cnum = \alpha \vnum$ clauses uniformly and with replacement from the
set of all $k$-clauses on $\vnum$ variables.  In terms of the factor
graph representation, this procedure samples a random
$(\vnum,\cnum)$-bipartite graph, in which each clause $a \in
\clauseset$ has degree $k$.

\paragraph{Markov random fields and marginalization:}  The $k$-SAT 
problem can also be associated with a particular distribution defined
as a Markov random field.  Recall that a given instance of $k$-SAT can
be specified by the collection of clause functions $\{ \clfun{\acl}:
\; \acl \in \clauseset\}$, as defined in
equation~\eqref{EqnDefnCompat}.  Using these functions, let us define
a probability distribution over binary sequences via
\begin{eqnarray}
\label{EqnSatDist}
p(\xvec) & \defn & \frac{1}{Z} \prod_{\acl \in \clauseset}
\clfun{\acl}(\xvec),
\end{eqnarray}
where $Z \defn \sum_{\xvec \in \{0,1\}^\vnum} \prod_{\acl \in
\clauseset} \clfun{\acl}(\xvec)$ is the normalization constant
%
Note that this definition makes sense if and only if the $k$-SAT
instance is satisfiable, in which case the
distribution~\eqref{EqnSatDist} is simply the uniform distribution
over satisfying assignments.
 
This Markov random field representation~\eqref{EqnSatDist} of any
satisfiable formula motivates a marginalization-based approach to
finding a satisfying assignment.  In particular, suppose that we had
an oracle that could compute exactly the marginal probability
\begin{equation}
\label{EqnDefnMarg}
p(x_i) = \sum_{\{x_j, j \in \vertexset \backslash \{i\} \}}
p(x_1, x_2, \ldots, x_n),
\end{equation}
for a particular variable $x_i$.  Note that this marginal reveals the
existence of SAT configurations with $x_i = 0$ (if $p(x_i =0) > 0$) or
$x_i = 1$ (if $p(x_i=1) > 0$).  Therefore, a SAT configuration could
be obtained by a recursive marginalization-decimation procedure,
consisting of computing the marginal $p(x_i)$, appropriately setting
$x_i$ (i.e., decimating), and then re-iterating the modified Markov
random field.

Of course, exact marginalization is computationally intractable in
general~\cite{Cooper90,Dagum93}, which motivates the use of efficient
algorithms for approximate marginalization.  An example of such an
algorithm is what we will refer to as the ``naive belief propagation
algorithm''.  The belief propagation (BP) algorithm, described in
detail in Appendix~\ref{AppBP}, can be applied to a MRF of the
form~\ref{EqnSatDist} to estimate the marginal probabilities. Even
though the BP algorithm is not exact, an intuitively reasonable
approach is to set the variable that has the largest bias towards a
particular value, and repeat. In fact, this marginalization-decimation
approach based on naive BP finds a satisfying assignment for $\alpha$
up to approximately 3.9 for $k=3$; for higher $\alpha$, however, the
iterations for BP typically fail to
converge~\cite{MPZ02,Aurell05,BMZ03}.

\vtiny 
 
\subsection{Survey propagation}

\label{subsec:sp}
In contrast to the naive BP approach, a marginalization-decimation
approach based on \emph{survey propagation} appears to be effective in
solving random $k$-SAT problems even close to
threshold~\cite{MPZ02,BMZ03}.  Here we provide an explicit description
of what we refer to as the $\SP(\rho)$ family of algorithms, where
setting the parameter $\rho = 1$ yields the pure form of survey
propagation.  For any given $\rho \in [0,1]$, the algorithm involves
updating messages from clauses to variables, as well as from variables
to clauses.  Each clause $\acl \in \clauseset$ passes a real number
$\messtovar{\acl}{i} \in [0,1]$ to each of its variable neighbors $i
\in \Nbrcl(\acl)$.  In the other direction, each variable $i \in
\vertexset$ passes a triplet of real numbers $\messtocl{i}{\acl} =
(\messtoclun{i}{\acl}, \messtoclsat{i}{\acl}, \messtoclstar{i}{\acl})$
to each of its clause neighbors $\acl \in \Nbrvar(i)$.  The precise
form of the updates are given in Figure~\ref{FigSPUpdates}.
\begin{figure}[h]
\framebox[.97\textwidth]{ 
\parbox{.88\textwidth}{
\vtiny {\emph{Message from clause $\acl$ to variable $i$:}}
\begin{equation}
\label{EqnUpMesstovar}
\messtovar{\acl}{i} = \prod_{j \in \Nbrcl(\acl) \backslash \{i\}}
\Biggl[ \frac{\messtoclun{j}{\acl}}{ \messtoclun{j}{\acl} +
\messtoclsat{j}{\acl} + \messtoclstar{j}{\acl}} \Biggr].
\end{equation}

{\emph{Message from variable $i$ to clause $\acl$:}}
\begin{subequations}
\label{EqnUpMesstocl}
\begin{eqnarray}
\label{EqnUpMesstoclun}
\messtoclun{i}{\acl} &=& \Biggr[ 1 -\rho \prod_{\bcl \in
\Nbrdis{\acl}{i}} (1-\messtovar{\bcl}{i})\Biggr] \prod_{b \in
\Nbragree{\acl}{i}} (1-\messtovar{\bcl}{i}). \\
\label{EqnUpMesstoclsat}
\messtoclsat{i}{\acl} &=& \Biggl[ 1 -\prod_{\bcl \in
\Nbragree{\acl}{i}} (1-\messtovar{\bcl}{i}) \Biggr] \prod_{b \in
\Nbrdis{\acl}{i}}(1-\messtovar{\bcl}{i}). \\
\label{EqnUpMesstoclstar}
\messtoclstar{i}{\acl} &=& \prod_{b\in \Nbragree{\acl}{i}} (1-
\messtovar{\bcl}{i} ) \prod_{b\in \Nbrdis{\acl}{i}} ( 1 -
\messtovar{b}{i} ).
\end{eqnarray} 
\end{subequations}
}} 
\caption{$\SP(\rho)$ updates}
\label{FigSPUpdates}
\end{figure}

We pause to make a few comments about these $\SP(\rho)$ updates:
\begin{enumerate}
\item 
Although we have omitted the time step index for simplicity,
equations~\eqref{EqnUpMesstovar} and~\eqref{EqnUpMesstocl} should be
interpreted as defining a recursion on $(\Messtovar, \Messtocl)$. The
initial values for $\Messtovar$ are chosen randomly in the interval
$(0, 1)$. 
\item The idea of the $\rho$ parameter is to provide a smooth
transition from the original naive belief propagation algorithm to the
survey propagation algorithm.  As shown in \cite{BMZ03}, setting $\rho
= 0$ yields the belief propagation updates applied to the probability
distribution~\eqref{EqnSatDist}, whereas setting $\rho=1$ yields the
pure version of survey propagation.
\end{enumerate}

\vtiny

\subsubsection{Intuitive ``warning'' interpretation}
\label{subsec:intuitive_sp}

To gain intuition for these updates, it is helpful to consider the
pure $\SP$ setting of $\rho = 1$.  As described by Braunstein et
al.~\cite{BMZ03}, the messages in this case have a natural
interpretation in terms of probabilities of warnings.  In particular,
at time $t=0$, suppose that the clause $\acl$ sends a warning message
to variable $i$ with probability $\Messtovar^{0}_{\acl \ra i}$, and a
message without a warning with probability $1-\Messtovar^0_{\acl \ra
i}$.  After receiving all messages from clauses in $\Nbrvar(i)
\backslash \{\acl\}$, variable $i$ sends a particular symbol to clause
$a$ saying either that it can't satisfy it (``u''), that it can
satisfy it (``s''), or that it is indifferent (``$*$''), depending on
what messages it got from its other clauses. There are four cases:
\begin{enumerate}
\item If variable $i$ receives warnings from $\Nbrdis{\acl}{i}$ and no
warnings from $\Nbragree{\acl}{i}$, then it cannot satisfy $\acl$ and
sends ``u''.
\item If variable $i$ receives warnings from $\Nbragree{\acl}{i}$ but
no warnings from $\Nbrdis{\acl}{i}$, then it sends an ``s'' to
indicate that it is inclined to satisfy the clause $\acl$.
\item If variable $i$ receives no warnings from either $\Nbrdis{\acl}{i}$ or
$\Nbragree{\acl}{i}$, then it is indifferent and sends ``$*$''.
\item If variable $i$ receives warnings from both $\Nbrdis{\acl}{i}$
and $\Nbragree{\acl}{i}$, a contradiction has occurred.
\end{enumerate}
The updates from clauses to variables are especially simple: in
particular, any given clause sends a warning if and only if it
receives ``u'' symbols from all of its other variables.

In this context, the real-valued messages involved in the pure
$\SP(1)$ all have natural probabilistic interpretations.  In
particular, the message $\messtovar{\acl}{i}$ corresponds to the
probability that clause $\acl$ sends a warning to variable $i$.  The
quantity $\messtoclun{j}{\acl}$ can be interpreted as the probability
that variable $j$ sends the ``u'' symbol to clause $\acl$, and
similarly for $\messtoclsat{j}{\acl}$ and
$\messtoclstar{j}{\acl}$. The normalization by the sum
$\messtoclun{j}{\acl} + \messtoclsat{j}{\acl} +
\messtoclstar{j}{\acl}$ reflects the fact that the fourth case is a
failure, and hence is excluded a priori from the probability
distribution 

Suppose that all of the possible warning events were independent.  In
this case, the SP message update equations~\eqref{EqnUpMesstovar}
and~\eqref{EqnUpMesstocl} would be the correct estimates for the 
probabilities. This independence
assumption is valid on a graph without cycles, and in that case the SP
updates do have a rigorous probabilistic interpretation.  It is not
clear if the equations have a simple interpretation in the case $\rho
\neq 1$.

\vtiny

\subsubsection{Decimation based on survey propagation}

Supposing that these survey propagation updates are applied and
converge, the overall conviction of a value at a given variable is
computed from the incoming set of equilibrium messages as
\comment{ 
\begin{equation}
\spmarg_i(1) \propto \Biggr[ 1 - \rho \prod_{\bcl \in \Nbrvarpos(j)}
(1-\messtovar{\bcl}{j})\Biggr] \prod_{b\in \Nbrvarneg(j)}(1-\eta_{\bcl
\ra j}),
\end{equation}
with similar formula for $\spmarg_i(0)$ and $\spmarg_i(*)$.
} 
\begin{eqnarray*}
\spmarg_i(1) & \propto & \Biggr[ 1 - \rho \prod_{\bcl \in
\Nbrvarpos(j)} (1-\messtovar{\bcl}{j})\Biggr] \prod_{b\in
\Nbrvarneg(j)}(1-\eta_{\bcl \ra j}).  \\
\spmarg_i(0) & \propto & \Biggr[ 1 - \rho \prod_{\bcl \in
\Nbrvarneg(j)} (1-\messtovar{\bcl}{j})\Biggr] \prod_{\bcl \in
\Nbrvarpos(j)}(1-\eta_{\bcl \ra j}). \\
\spmarg_i(*) & \propto & \prod_{b\in
\Nbrvarpos(j)}(1-\messtovar{\bcl}{j}) \prod_{\bcl \in
\Nbrvarneg(j)}(1-\messtovar{\bcl}{j}). \\
\end{eqnarray*} 
To be consistent with their interpretation as (approximate) marginals,
the triplet $\{\spmarg_i(0), \spmarg_i(*), \spmarg_i(1) \}$ at each
node $i \in \vertexset$ is normalized to sum to one.  We define the
\emph{bias} of a variable node as \mbox{$\Bias(i) \defn |\spmarg_i(0)-
\spmarg_i(1)|$.}  

The marginalization-decimation algorithm based on survey
propagation~\cite{BMZ03} consists of the following steps:
\begin{enumerate}
\item Run $\SP(1)$ on the SAT problem.  Extract the fraction
$\biasfrac$ of variables with the largest biases, and set them to
their preferred values.
\item Simplify the SAT formula, and return to Step 1.
\end{enumerate}
Once the maximum bias over all variables falls below a pre-specified
tolerance, the Walk-SAT algorithm is applied to the formula to find
the remainder of the assignment (if possible).
Intuitively, the goal of the initial phases of decimation is to find a
cluster; once inside the cluster, the induced problem is considered easy to
solve, meaning that any ``local'' algorithm should perform well within a
given cluster. 

\vtiny

\section{Markov random fields over partial assignments}
\label{sec:MRF}

In this section, we show how a large class of message-passing
algorithms---including the $\SP(\rho)$ family as a particular
case---can be recovered by applying the well-known belief propagation
algorithm to a novel class of Markov random fields (MRFs) associated
with any $k$-SAT problem.  We begin by introducing the notion of a
partial assignment, and then use it to define the family of MRFs over
these assignments.

\vtiny

\subsection{Partial assignments} 

Suppose that the variables $x = \{x_1, \ldots, x_n\}$ are allowed to
take values in $\{0, 1, *\}$, which we refer to as a \emph{partial
assignment}.  It will be convenient, when discussing the assignment of
a variable $x_i$ with respect to a particular clause $\acl$, to use
the notation $\sat{\acl}{i} := 1 - \Jsca_{\acl,i}$ and
$\unsat{\acl}{i} \; \defn \; \Jsca_{\acl,i}$ to indicate,
respectively, the values that are \emph{satisfying} and
\emph{unsatisfying} for the clause $\acl$.  With this set-up, we have
the following:
\begin{definition}
A partial assignment $\xvec$ is \emph{invalid} for a clause $\acl$
if either
\begin{enumerate}
\item[(a)] all variables are unsatisfying (i.e.,
$\xvar_i=\unsat{\acl}{i}$ for all $i \in \Nbrcl(\acl)$), or
\item[(b)] all variables are unsatisfying except for exactly one index
$j \in \Nbrcl(\acl)$, for which $x_j=*$.
\end{enumerate}
Otherwise, the partial assignment $\xvec$ is valid for clause
$\acl$, and we denote this event by $\Val{\acl}{\xvar_{\Nbrcl(\acl)}}$.
We say that a partial assignment is \emph{valid} for a
formula if it is valid for all of its clauses.
\end{definition}

The motivation for deeming case (a) invalid is clear, in
that any partial assignment that does not satisfy the clause must
be excluded.  Note that case (b) is also invalid, since (with all
other variables unsatisfying) the variable $x_j$ is effectively forced
to $\sat{\acl}{i}$, and so cannot be assigned the $*$ symbol.

For a valid partial assignment, the subset of variables that are
assigned either 0 or 1 values can be divided into \emph{constrained}
and \emph{unconstrained} variables in the following way:
\begin{definition}
We say that a variable $x_i$ is the \emph{unique satisfying variable}
for a clause if it is assigned $\sat{\acl}{i}$ whereas all other
variables in the clause (i.e., the variables $\{ \xvar_j \; : \; j \in
\Nbrcl(\acl)\backslash \{i\} \}$) are assigned $\unsat{\acl}{j}$.  A
variable $x_i$ is \emph{constrained} by clause $a$ if it is the unique
satisfying variable.
\end{definition}
We let $\Con{i}{\acl}(\xvar_{\Nbrcl(\acl)})$ denote an indicator
function for the event that $\xvar_i$ is the unique satisfying
variable in the partial assignment $\xvec_{\Nbrcl(\acl)}$ for clause
$\acl$. A variable is \emph{unconstrained} if it has 0 or 1 value, and
is not constrained. Thus for any partial assignment the variables are
divided into stars, constrained and unconstrained variables.  We
define the three sets
\begin{equation}
\label{EqnDefnSets}
\begin{array}{lll}
\Setstar(\xvec) := \{i \in \vertexset: x_i =*\} &
\Setcon(\xvec) := \{i \in \vertexset: x_i \mbox{$\;$ constrained} \} &
\Setunc(\xvec) := \{i \in \vertexset: x_i \mbox{$\;$ unconstrained} \} 
\end{array}
\end{equation}
of $\ast$, constrained and unconstrained variables respectively.
Finally, we use $n_*(x)$, $n_c(x)$ and $n_\unc(x)$ to denote the
respective sizes of these three sets.

Various probability distributions can be defined on valid partial
assignments by giving different weights to stars, constrained and
unconstrained variables, which we denote by $\wcon$, $\wstar$ and
$\wun$ respectively.  Since only the ratio of the weights matters, we
set $\wcon=1$, and treat $\wun$ and $\wstar$ as free non-negative
parameters (we generally take them in the interval $[0,1]$).  We
define the weights of partial assignments in the following way:
invalid assignments $\xvec$ have weight $\Weight(\xvec)=0$, and for
any valid assignment $\xvec$, we set
\begin{equation}
\label{EqnDefnWeight}
\Weight(\xvec) \defn (\wun)^{n_\unc(\xvec)} \times
(\wstar)^{n_*(\xvec)}.
\end{equation}
Our primary interest is the probability distribution given by
$p_\Weight(\xvec) \propto \Weight(\xvec)$.  In contrast to the earlier
distribution $p$, it is important to observe that this definition is
valid for any SAT problem, whether or not it is satisfiable, as long
as $\wstar \neq 0$, since the all-$*$ vector is always a valid
partial assignment. Note that if $\wun = 1$ and $\wstar = 0$ then
the distribution $p_\Weight(\xvec)$ is the uniform distribution on
satisfying assignments. Another interesting case that we will discuss is 
that of $\wun=0$ and $\wstar=1$, which corresponds to the uniform 
distribution over valid partial assignments without unconstrained 
variables.

\vtiny

\subsection{Associated Markov random fields}  

Given our set-up thus far, it is not at all obvious whether or not the
distribution $p_\Weight$ can be decomposed as a Markov random field
based on the original factor graph.  Interestingly, we find that
$p_\Weight$ does indeed have such a Markov representation for any
choices of $\wun, \wstar \in [0,1]$.  Obtaining this representation
requires the addition of another dimension to our representation,
which allows us to assess whether a given variable is constrained or
unconstrained. We define the \emph{parent set} of a given variable
$x_i$, denoted by $\Par{i}$, to be the set of clauses for which $x_i$
is the unique satisfying variable.
Immediate consequences of this definition are the following:
\begin{enumerate}
\item[(a)] If $x_i=0$, then we must have $\Par{i} \subseteq
\Nbrvarzero(i)$.  
\item[(b)] If $x_i=1$, then there must hold $\Par{i} \subseteq
\Nbrvarone(i)$.
\item[(c)] The setting $x_i=*$ implies that $\Par{i}=\emptyset$.
\end{enumerate}
Note also that $\Par{i} = \emptyset$ means that $x_i$ cannot be
constrained.  
For each $i \in \vertexset$, let $\Suppar(i)$ be the set of all
possible parent sets of variable $i$.  Due to the restrictions imposed
by our definition, $\Par{i}$ must be contained in either
$\Nbrvarone(i)$ or $\Nbrvarzero(i)$ but not both. Therefore, the
cardinality\footnote{Note that it is necessary to subtract one so as
not to count the empty set twice.}  of $\Suppar(i)$ is
$2^{|\Nbrvarzero(i)|} + 2^{|\Nbrvarone(i)|} -1$.

Our extended Markov random field is defined on the Cartesian product
space $\extX_1 \times \ldots \times \extX_\vnum$, where $\extX_i \defn
\{0,1,*\} \times \Suppar(i)$.  The distribution factorizes as a
product of compatibility functions at the variable and clause nodes of
the factor graph, which are defined as follows:

\paragraph{Variable compatibilities:}  Each variable node $i \in \vertexset$
has an associated compatibility function of the form:
\begin{eqnarray}
\label{EqnDefnNodeSupcompat}
\supcompat_i(x_i, \Par{i}) & \defn & \left\{\begin{array}
{r@{\quad:\quad}l} \wun & \Par{i} = \emptyset, \xvar_i\neq * \\
\wstar & \Par{i} = \emptyset, \xvar_i=* \\
1 & \mbox{for any other valid $(\Par{i}, \xvar_i)$}
\end{array} \right.
\end{eqnarray}
The role of these functions is to assign weight to the partial
assignments according to the number of unconstrained and star
variables, as in the weighted distribution $p_{\Weight}$.

\paragraph{Clause compatibilities:} The compatibility functions at the
clause nodes serve to ensure that only valid assignments have non-zero
probability, and that the parent sets \mbox{$\Par{\Nbrcl(\acl)} \defn
\{ \Par{i} \; : \; i \in \Nbrcl(\acl) \}$} are consistent with the
assignments $\xvar_{\Nbrcl(\acl)} \defn \{ \xvar_i \; : \; i \in
\Nbrcl(\acl) \}$ in the neighborhood of $\acl$.  More precisely, we
require that the partial assignment $\xvar_{\Nbrcl(\acl)}$ is
valid for $\acl$ (i.e., $\Val{\acl}{\xvar_{\Nbrcl(\acl)}} = 1$) and
that for each $i \in \Nbrcl(\acl)$, exactly one of the two following
conditions holds:
\begin{enumerate}
\item[(a)] $\acl \in \Par{i}$ and $x_i$ is constrained by $\acl$
or
\item[(b)] $\acl \notin \Par{i}$ and $x_i$ is not constrained by
$\acl$.
\end{enumerate} 

The following compatibility function corresponds to an indicator
function for the intersection of these events:
\begin{equation}
\label{EqnDefnClSupcompat}
\supcompat_{\acl} \big( \xvar_{\Nbrcl(\acl)}, \Par{\Nbrcl(\acl)} \big)
:= \Val{\acl}{\xvar_{\Nbrcl(\acl)}} \times \prod_{i \in \Nbrcl(\acl)}
\delta \bigl( \Ind{\acl \in \Par{i}}, \; \;
\Con{\acl}{i}(\xvar_{\Nbrcl(\acl)}) \bigr).
\end{equation}
We now form a Markov random field over partial assignments and
parent sets by taking the product of
variable~\eqref{EqnDefnNodeSupcompat} and
clause~\eqref{EqnDefnClSupcompat} compatibility functions
\begin{equation}
\label{EqnDefnSuperMRF}
\superp(\xvec, \Par{}) \propto  \prod_{i \in \vertexset}
\supcompat_i(\xvar_i, \Par{i}) \; \; \prod_{\acl \in \clauseset }
\supcompat_\acl \big ( \xvar_{\Nbrcl_\acl}, \Par{\Nbrcl(\acl)} \big).
\end{equation}
With these definitions, some straightforward calculations show that
$\superp = p_{\Weight}$.

\vtiny

\subsection{Survey propagation as an instance of belief propagation}
\label{SecSPisBP}

We now consider the form of the belief propagation (BP) updates as
applied to the MRF $\superp$ defined by
equation~\eqref{EqnDefnSuperMRF}.  We refer the reader to Section
\ref{AppBP} for the definition of the BP algorithm on a general factor
graph.  The main result of this section is to establish that the
$\SP(\rho)$ family of algorithms are equivalent to belief propagation
as applied to $\superp$ with suitable choices of the weights $\wun$
and $\wstar$. In the interests of readability, most of the technical
lemmas will be presented in the appendix.

We begin by introducing some notation necessary to
describe the BP updates on the extended MRF. The BP message from
clause $\acl$ to variable $i$, denoted by $\Genmess_{\acl \ra
i}(\cdot)$, is a vector of length $|\extX_i| = 3 \times |\Suppar(i)|$.
Fortunately, due to symmetries in the variable and clause
compatibilities defined in equations~\eqref{EqnDefnNodeSupcompat}
and~\eqref{EqnDefnClSupcompat}, it turns out that the
clause-to-variable message can be parameterized by only three numbers,
$\{\tovarun{\acl}{i}, \tovarsat{\acl}{i}, \tovarstar{\acl}{i} \}$, as
follows:
\begin{eqnarray}
\Genmess_{\acl \ra i}(x_i, \Par{i} ) & = &
\begin{cases} \tovarsat{\acl}{i}  & \mbox{ if } \xvar_i = \sat{\acl}{i}, \;  
\Par{i} = \sset \cup \{\acl\} \mbox{ for some } \sset \subseteq
\Nbragree{\acl}{i}, \\
\tovarun{\acl}{i} & \mbox{ if } \xvar_i = \unsat{\acl}{i}, \;\Par{i}
\subseteq \Nbrdis{\acl}{i}, \\
\tovarstar{\acl}{i} & \mbox{ if } \xvar_i = \sat{\acl}{i}, \; \Par{i}
\subseteq \Nbragree{\acl}{i} \mbox{ or } \xvar_i = * ~, \Par{i} =
\emptyset,\\
0 & \mbox{ otherwise.}
\end{cases}
\end{eqnarray}
where $\tovarsat{\acl}{i}, \tovarun{\acl}{i}$ and
$\tovarstar{\acl}{i}$ are elements of $[0, 1]$.

Now turning to messages from variables to clauses, it is convenient to
introduce the notation $\Par{i} = S \cup \{a\}$ as a shorthand for the
event
\begin{eqnarray*}
\acl \in \Par{i} & \mbox{ and } & S = \Par{i}\backslash \{\acl\}
\subseteq \Nbragree{\acl}{i},
\end{eqnarray*}
where it is understood that $S$ could be empty.  In
Appendix~\ref{AppExtBP}, we show that the variable-to-clause message
$\Toclmess_{i \ra \acl}$ is fully specified by values for pairs $(x_i,
\Par{i})$ of six general types:
\[
\{ 
(\sat{\acl}{i}, S \cup \{a\}), \; \; 
(\sat{\acl}{i}, \emptyset \neq \Par{i} \subseteq \Nbragree{\acl}{i}), \; \; 
(\unsat{\acl}{i}, \emptyset \neq \Par{i} \subseteq \Nbrdis{\acl}{i}), \; \; 
(\sat{\acl}{i}, \emptyset), \; \; 
(\unsat{\acl}{i}, \emptyset), \; \; 
(*, \emptyset) \}.
\]
The BP updates themselves are most compactly expressed in terms of
particular linear combinations of such basic messages, defined in the
following way:
\begin{subequations}
\label{EqnDefnR}
\begin{eqnarray}
\label{EqnDefnRsat}
\Rsat{i}{\acl} & \defn & \sum_{S \subseteq \Nbragree{\acl}{i}}
\Toclmess_{i \ra \acl}(\sat{\acl}{i}, S \cup \{a \}) \\
\label{EqnDefnRun}
\Run{i}{\acl} & \defn & \sum_{\Par{i} \subseteq \Nbrdis{\acl}{i}}
\Toclmess_{i \ra a}(\unsat{\acl}{i}, \; \Par{i}) \\
\label{EqnDefnRstar}
\Rstar{i}{\acl} & \defn & \sum_{ \Par{i} \subseteq \Nbragree{\acl}{i}} 
\Toclmess_{i \ra a}(\sat{\acl}{i}, \Par{i}) + 
\Toclmess_{i \ra a}(*, \emptyset).
\end{eqnarray}
\end{subequations}
Note that $\Rsat{i}{\acl}$ is associated with the event that $x_i$ is
the unique satisfying variable for clause $\acl$; $\Run{i}{\acl}$ with
the event that $x_i$ does not satisfy $\acl$; and $\Rstar{i}{\acl}$
with the event that $x_i$ is neither unsatisfying nor uniquely
satisfying (i.e., either $x_i = *$, or $x_i = \sat{\acl}{i}$ but is
not the only variable that satisfies $\acl$).

With this terminology, the BP algorithm on the extended MRF can be
expressed in terms of the following recursions on the triplets
$(\tovarsat{\acl}{i}, \tovarun{\acl}{i}, \tovarstar{\acl}{i})$ and
$(\Rsat{i}{\acl}, \Run{i}{\acl}, \Rstar{i}{\acl} )$: \\

\vspace*{.1in}
\framebox[.98\textwidth]{ 
\parbox{.95\textwidth}{
\vtiny 
{\underline{BP updates on extended MRF:}} \\
{\small{ \emph{Messages from clause $\acl$ to variable $i$}
\begin{subequations}
\begin{eqnarray*}
\tovarsat{\acl}{i} &=& \prod_{j\in \Nbrvar(\acl) \backslash \{i\}}
 \Run{j}{\acl} \\
\tovarun{\acl}{i} & = & \prod_{j\in \Nbrvar(\acl) \backslash \{i\}}
(\Run{j}{\acl} + \Rstar{j}{\acl}) + \sum_{k \in \Nbrvar(\acl)
\backslash \{i\}} (\Rsat{k}{\acl} - \Rstar{k}{\acl}) \prod_{j\in
\Nbrvar(\acl) \backslash \{i, k\}} \Run{j}{\acl} - \prod_{j \in
\Nbrvar(\acl) \backslash \{i\}} \Run{j}{\acl}  \qquad \qquad \\
\tovarstar{\acl}{i} & = & \prod_{j\in \Nbrvar(\acl) \backslash \{i\}}
(\Run{j}{\acl} + \Rstar{j}{\acl})- \prod_{j\in \Nbrvar(\acl)
\backslash \{i\}} \Run{j}{\acl}.
\end{eqnarray*}
\end{subequations}
\emph{Messages from variable $i$ to clause $\acl$:}
\begin{subequations}
\begin{eqnarray*}
\Rsat{i}{\acl} & = & \prod_{b \in \Nbrdis{\acl}{i}} \tovarun{b}{i}
\Bigl[ \prod_{b \in \Nbragree{\acl}{i}} (\tovarsat{b}{i} + \tovarstar{b}{i}) \Bigr]
\\
\Run{i}{\acl} &=& \prod_{b\in \Nbragree{\acl}{i}} \tovarun{b}{i}
\Bigl[ \prod_{b \in \Nbrdis{\acl}{i}}(\tovarsat{b}{i} +
\tovarstar{b}{i}) - (1-\wun) \prod_{b \in V^u_a(i)} \tovarstar{b}{i}
\Bigr] \\
\Rstar{i}{\acl} & = & \prod_{b \in \Nbrdis{\acl}{i}} \tovarun{b}{i}
\Bigl[ \prod_{b \in \Nbragree{\acl}{i}}(\tovarsat{b}{i} +
\tovarstar{b}{i}) - (1-\wun) \prod_{b \in \Nbragree{\acl}{i}}
\tovarstar{b}{i} \Bigr] + \wstar \prod_{b \in \Nbragree{\acl}{i} \cup
\Nbrdis{\acl}{i} } \tovarstar{b}{i}.
\end{eqnarray*}
\end{subequations}
}}
}}

\vspace*{.1in}

We provide a detailed derivation of these BP equations on the extended
MRF in Appendix~\ref{AppExtBP}.  Since the messages are interpreted as
probabilities, we only need their ratio, and we can normalize them to
any constant. At any iteration, approximations to the local marginals
at each variable node $i \in \vertexset$ are given by (up to a
normalization constant):
\begin{eqnarray*}
\Field_i(0) & \propto & \prod_{b\in \Nbrvarpos(i)} \tovarun{b}{i}
\Bigl[ \prod_{b \in \Nbrvarneg(i)} (\tovarsat{b}{i} +
\tovarstar{b}{i}) - (1-\wun) \prod_{b \in \Nbrvarneg(i)}
\tovarstar{b}{i} \Bigr] \\
\Field_i(1) & \propto & \prod_{b\in \Nbrvarneg(i)} \tovarun{b}{i} \Bigl[
\prod_{b \in \Nbrvarpos(i)} (\tovarsat{b}{i} + \tovarstar{b}{i}) -
(1-\wun) \prod_{b \in \Nbrvarpos(i)} \tovarstar{b}{i} \Bigr] \\
\Field_i(*) & \propto & \wstar \prod_{b \in \Nbrvar(i)} \tovarstar{b}{i}
\end{eqnarray*}

The following theorem establishes that the $\SP(\rho)$ family of
algorithms is equivalent to belief propagation on the extended MRF:

\begin{theorem} \label{thm:main}
For all $\wstar \in [0,1]$, the BP updates on the extended $(\wstar,
\wun)$-MRF $\superp$ are equivalent to the $\SP(\wstar)$ family of
algorithms under the following restrictions:
\begin{enumerate}
\item[(a)] the constraint $\wun + \wstar = 1$ is imposed, and
\item[(b)] 
all messages are initialized such that $\tovarun{\acl}{i}=
\tovarstar{\acl}{i}$ for every edge $(\acl,i)$.
\end{enumerate}
\begin{proof}
Under the constraint $\wun + \wstar = 1$, if we initialize
$\tovarun{\acl}{i}= \tovarstar{\acl}{i}$ on every edge, then there
holds $\Rsat{i}{\acl} = \Rstar{i}{\acl}$ and consequently
$\tovarun{\acl}{i} = \tovarstar{\acl}{i}$ remains true at the next
iteration.  Initializing the parameters in this way and imposing the
normalization $\tovarun{\acl}{i} + \tovarstar{\acl}{i} = 1$ leads to
the following recurrence equations:
\begin{equation*}
\tovarsat{a}{i} = \frac {\prod_{j \in \Nbrvar(\acl) \backslash\{i\}}
\Run{j}{\acl} } {\prod_{j \in \Nbrvar(\acl) \backslash\{i\}}
(\Rstar{j}{\acl} + \Run{j}{\acl} )}
\end{equation*}
where:
\begin{eqnarray*}
\Run{i}{\acl} & = & \prod_{b\in \Nbragree{\acl}{i}} (1-
\tovarstar{b}{i} ) \Bigl[ 1 - \wstar \prod_{b\in \Nbrdis{\acl}{i}} (1
- \tovarstar{b}{i}) \Bigr] \\
\Rstar{i}{\acl} & = & \prod_{b\in \Nbrdis{a}{i}} (1-\tovarsat{b}{i}).
\end{eqnarray*}
These updates are equivalent to $\SP(\wstar)$ by setting $\eta_{\acl
\ra i} = \tovarsat{\acl}{i}$, $\Pi_{i \ra \acl}^u = \Run{i}{\acl}$,
and $\Pi_{i \ra \acl}^s + \Pi_{i \ra \acl}^* = \Rstar{i}{\acl}$.
\end{proof}
\end{theorem}

\paragraph{Remarks:}
\begin{enumerate}
\item Theorem~\ref{thm:main} is a generalization of the result of
Braunstein and Zecchina~\cite{BZ04}, who showed that $\SP(1)$ is
equivalent to belief propagation on a certain MRF.

\item The essence of Theorem~\ref{thm:main} is that the pure survey
propagation algorithm, as well as all the $\rho$-variants thereof, are
all equivalent to belief propagation on our extended MRF with suitable
parameter choices. This equivalence is important for a number of
reasons:
\begin{enumerate}
\item Belief propagation is a widely-used algorithm for computing
approximations to marginal distributions in general Markov random
fields~\cite{YFW03,Frank01}.  It also has a variational interpretation
as an iterative method for attempting to solve a non-convex
optimization problem based on the Bethe approximation~\cite{YFW03}.
Among other consequences, this variational interpretation leads to
other algorithms that also solve the Bethe problem, but unlike belief
propagation, are guaranteed to
converge~\cite{Welling01,Yuille01,WJ03}.

\comment{Conditions for convergence of the algorithm were also studied
in a general setting \cite{TJ02}.}

\item Given the link between $\SP$ and extended MRFs, it is natural to
study combinatorial and probabilistic properties of the latter
objects.  In Section~\ref{SecCore}, we show how so-called``cores''
arise as fixed points of $\SP(1)$, and we prove a weight-preserving
identity that shows how the extended MRF for general $\rho$ is a
``smoothed'' version of the naive MRF.
\item Finally, since BP (and hence SP) is computing approximate
marginals for the MRF, it is natural to study other ways of computing
marginals and examine if these lead to an effective way for solving
random $k$-SAT problems. We begin this study in
Section~\ref{SecGibbs}.

\end{enumerate}



\item The initial messages have very small influence on the behavior
of the algorithm, and they are typically chosen to be uniform random
variables in $(0, 1)$. In practice, for $\wun + \wstar = 1$ if we
start with different values for $\tovarun{\acl}{i}$ and
$\tovarstar{\acl}{i}$ they soon converge to become equal.

\item If we restrict our attention to 3-SAT, the equations have
simpler form. In particular for a clause $\acl$ on $x_i$, $x_j$,
$x_k$, the messages to variable node $i$ are:
\begin{eqnarray*}
\tovarstar{a}{i} & = & \Run{j}{\acl} \Run{k}{\acl} \\
\tovarun{a}{i} & = & \Rstar{j}{\acl} \Rstar{k}{\acl} + \Rsat{j}{\acl}
\Run{k}{a} + \Run{j}{a} \Rsat{k}{\acl} \\
\tovarstar{a}{i} & = & \Rstar{j}{a} \Rstar{k}{a} + \Rstar{j}{a}
\Run{k}{\acl} + \Run{j}{a} \Rstar{k}{a}.
\end{eqnarray*}

\end{enumerate}

\vtiny

\section{Combinatorial properties} 
\label{SecCore}

This section is devoted to an investigation of the combinatorial
properties associated with the family of extended Markov random fields
defined in the previous section.  We begin by defining an acyclic directed
graph on all valid partial assignments. Of particular interest are the 
minimal elements in the resulting partial ordering. We refer to these 
as \emph{cores}.

\vtiny
\subsection{Directed graph and partial ordering}

The vertex set of the directed graph $\dirgraph$ consists of all valid
partial assignments.  The edge set is defined in the following
way: for a given pair of valid partial assignments $x$ and $y$,
the graph includes a directed edge from $x$ to $y$ if there exists an
index $i \in \vertexset$ such that (i) $x_j = y_j$ for all
$j \neq i$; and (ii) $y_i = \ast$ and $x_i \neq y_i$.  We label the
edge between $x$ and $y$ with the index $i$, corresponding to the fact
that $y$ is obtained from $x$ by adding one extra $\ast$ in position
$i$.

This directed graph $\dirgraph$ has a number of properties:
\begin{enumerate}
\item[(a)] Valid partial assignments can be separated into
 different levels based on their number of star variables.
 In particular, assignment $x$ is in level $n_*(x)$. Thus, every edge
 is from an assignment in level $l-1$ to one in level $l$, where $l$ is at
 most $n$.
\item[(b)] The out-degree of any valid partial assignment $x$ is
exactly equal to its number of unconstrained variables $n_\unc(x)$.
\item[(c)] It is an acyclic graph so that its structure defines a
partial ordering; in particular, we write $y < x$ if there is a
directed path in $\dirgraph$ from $x$ to $y$. Notice that all directed
paths from $x$ to $y$ are labeled by indices in the set $T=\{i \in
\vertexset: x_i \neq y_i=* \}$, and only the order in which they
appear is different.
\end{enumerate}

Given the partial ordering defined by $\dirgraph$, it is natural to
consider elements that are minimal in this partial ordering.  For any
valid partial assignment $x$ and a subset $S \subseteq
\vertexset$, let $\Core{S}{x}$ be the minimal $y < x$, such that the
path from $x$ to $y$ is labeled only by indices in $S$. In particular 
$\Core{V}{x}$ is a minimal assignment in the order. It is easy to
show that there always exists a unique $\Core{S}{x}$.  

\begin{proposition}
\label{prop:core}
For any valid assignment $x$ and $S \subseteq \vertexset$,
there is a unique minimal $y<x$ such that the path from $x$ to $y$ is
labeled only by indices in $S$. Furthermore $\Setunc(y)\cap S=\emptyset$
and $\Setstar(y) = \Setstar(x) \cup T$, where $T \subseteq S$ is the set of
labels on any path from $x$ to $y$.
\end{proposition}

\begin{proof}
To prove the second assertion in the proposition statement for a
minimal $y$, suppose there exists $i \in S \cap \Setunc(y)$. Then
there must be be an outgoing edge from $y$ labeled by an element in
$S$, which contradicts the assumed minimality of $y$. The equivalence
$\Setstar(y)=\Setstar(x) \cup T$ follows directly from the definition
of $\graph$ and its edge labels.

To establish the uniqueness statement, suppose that there are two
minimal such assignments $y_1$ and $y_2$, and the paths from $x$ to
$y_1$ and $y_2$ are labeled by sets of indices $T_1, T_2 \subseteq S$
respectively. If $T_1=T_2$ then $y_1=y_2$, so let us assume that $T_1$
and $T_2$ are distinct.  Without loss of generality, we may take $T_1
\backslash T_2 \neq \emptyset$. Consider a particular path from $x$ to
$y_1$, with labels $t_1, t_2, \ldots t_r$, where $r=|T_1|$. Let $t_i$
be the first label such that $t_i \notin T_2$. Then its corresponding
variable is unconstrained when the variables indexed by $ \{t_1,
\ldots t_{i-1}\} \cup \Setstar(x) \subseteq T_2 \cup \Setstar(x)$ are
assigned $*$, therefore it is unconstrained in $y_2$. This implies
that there exists an edge out of $y_2$ that is labeled by $t_i\in S$,
which contradicts the assumption that $y_2$ is minimal.
\end{proof}

We define a \emph{core assignment} to be a valid partial assignment $y
\in\{0,1,*\}^\vnum$ that contains no unconstrained variables.  We say
that a core assignment $y$ is \emph{non-trivial} if $n_*(y) < \vnum$,
so that it has at least one constrained $\{0,1\}$ variable.  Under
this definition, it follows that for any partial assignment $x$, the
associated minimal element $\Core{\vertexset}{x}$ is a core
assignment.

Given a valid ordinary assignment $z \in \{0,1\}^\vnum$, an
interesting object is the subgraph of partial assignments that lie
below it in the partial ordering.  It can be seen that any pair of
elements in this subgraph have both a unique maximal element and a
unique minimal element, so that any such subgraph is a
lattice~\cite{Stanley1}.

\begin{figure}
\begin{center}
\begin{tabular}{c}
\widgraph{.6\textwidth}{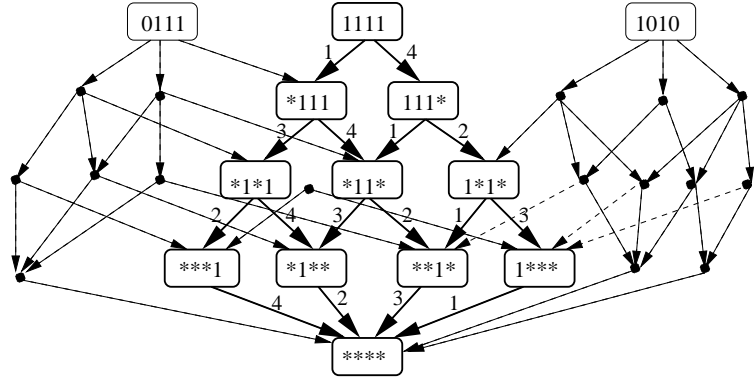}  \\
\\
(a) \\
\\
\widgraph{.5\textwidth}{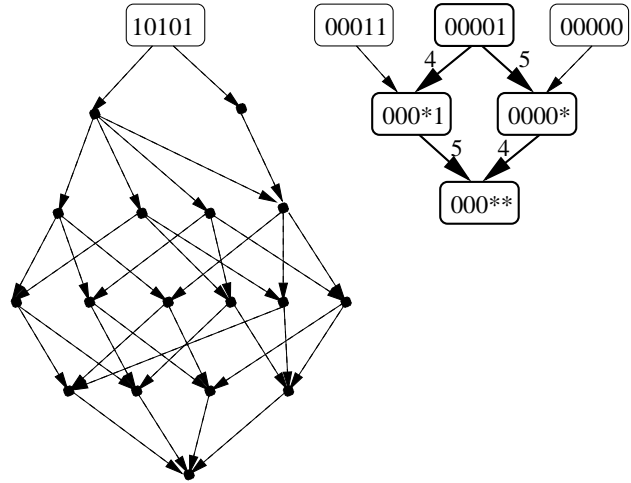}\\
\\
(b) \\
\\
\end{tabular}
\caption{ Portion of the directed graph on partial assignments for two
different formulas:
(a) $(\bar{x}_1 \vee \bar{x}_2 \vee x_3) \wedge 
( x_2 \vee \bar{x}_3 \vee \bar{x}_4)$. 
highlighted is the lattice below the satisfying assignment 
$z = (1,1,1,1)$, whose core is trivial (i.e.,  
$\Core{\vertexset}{z} = (\ast, \ast, \ast, \ast)$).
(b) $(\bar{x}_1 \vee x_2 \vee x_3) 
\wedge (x_1 \vee \bar{x}_2 \vee x_3) 
\wedge (x_2 \vee \bar{x}_3 \vee x_1) 
\wedge (x_2 \vee \bar{x}_3 \vee x_5) 
\wedge (x_1 \vee x_5 \vee \bar{x}_4)$.
the satisfying assignment $z= (0,0,0,0,1)$ has the
non-trivial core $\Core{\vertexset}{z} = (0,0,0,*,*)$. For the same 
formula there are other satisfying assignments, e.g. $(1,0,1,0,1)$
which have a trivial core. }
\label{fig:lattice}
\end{center}
\end{figure}

In examples shown in Figure~\ref{fig:lattice}, only a subset of the
partial assignments is shown, since even for small formulas the space
of partial assignments is quite large. For the first formula all
satisfying assignments have a trivial core. For the second one, on the
other hand, there are assignments with non-trivial cores.

\vtiny

\subsection{Pure survey propagation as a peeling procedure} 
\label{subsubsec:cores}

As a particular case of Theorem~\ref{thm:main}, setting $\wstar=1$ and
$\wun=0$ yields the extended MRF that underlies the $\SP(1)$
algorithm. In this case, the only valid assignments with positive
weight are those without any unconstrained variables---namely, core
assignments.  Thus, the distribution $p_{\Weight}$ for $(\wun, \wstar)
= (0,1)$ is simply uniform over the core assignments.  The following
result connects fixed points of $\SP(1)$ to core assignments:

\begin{proposition}
\label{prop:SPtocore}
For a valid assignment $x$, let $\SP(1)$ be initialized by:
\[
\begin{array}{lll}
\messtoclun{i}{\acl} = \delta(x_i, \unsat{\acl}{i}),&
\messtoclsat{i}{\acl} = \delta (x_i, \sat{\acl}{i}),&
\messtoclstar{i}{\acl} = 0.
\end{array}
\]
Then within a finite number of steps, the algorithm converges and the
output fields are
$$\spmarg_i(b) = \delta(y_i, b),$$ where $y=\Core{V}{x}$ and $b\in
\{0, 1, *\}$.  
\end{proposition}

\begin{proof}
We say that a variable $i$ belongs to the core if $y_i \neq *$.  We
say that a clause $\acl$ belongs to the core if all the variables in
the clause belong to the core.  We first show by induction that
\begin{enumerate}
\item[I.] If $\acl$ and $i$ belong to the core and $y_i$ is not the
unique satisfying variable for $\acl$ then $\messtoclun{i}{\acl} =
\delta(x_i, \unsat{\acl}{i})$ and $\messtoclsat{i}{\acl} = \delta
(x_i, \sat{\acl}{i})$, and
\item[II.] If $\acl$ and $i$ belong to the core and $y_i$ is the
unique satisfying variable for $\acl$ then $\messtovar{\acl}{i} = 1$.
\end{enumerate}
Clearly, property I holds at time $0$. Therefore, it suffices to prove
that if property I holds at time $t$ then so does II. and that if
property II holds at time $t$ then property I holds at time $t+1$.

Suppose that property I holds at time $t$.  Let $\acl$ and $i$ belong
to the core such that $y_i$ is the unique satisfying variable of the
clause $\acl$. By the induction hypothesis for all $j \in
V(a)\setminus\{i\}$ it holds that $\messtoclun{j}{\acl} = \delta(x_j,
\unsat{\acl}{j}) = 1$. This implies that $\messtovar{\acl}{i} = 1$ as
needed.

Suppose that property II holds at time $t$. Let $\acl$ and $i$ belong
to the core such that $y_i$ is not unique satisfying for $\acl$. By
the assumption, it follows that there exists $\bcl$ which belongs to
the core such that $y_i$ is the unique satisfying variable for
$\bcl$. This implies by the induction hypothesis that
$\messtovar{\bcl}{i} = 1$. It is now easy to see that at update $t+1$:
$\messtoclun{i}{\acl} = \delta(x_i, \unsat{\acl}{i})$ and
$\messtoclsat{i}{\acl} = \delta (x_i, \sat{\acl}{i})$.  Note that the
claim above implies that for all times $t$ and all $i$ such that $y_i
\neq *$, it holds that $\spmarg_i(b) = \delta(y_i, b)$.

Let $i_1,i_2,\ldots,i_s$ be a ``peeling-path'' from $x$ to $y$. In
other words, the variable $i_1$ is not uniquely satisfying any
clause. Once, this variable is set to $*$, the variable $i_2$ is not
uniquely satisfying any clause etc. We claim that for all $1 \leq t
\leq s$, for all updates after time $t$ and for all clauses $\acl$
such that $i_t \in V(a)$ it holds that $\messtovar{\acl}{i_t} =
0$. The proof follows easily by induction on $t$.  This in turn
implies that if for all updates after time $t$ $\spmarg_{i_t}(b) =
\delta(y_i,*)$, from which the result follows.
\end{proof}

Thus, $\SP(1)$, when suitably initialized, simply strips the valid
assignment $x$ down to its core $\Core{V}{x}$.  Moreover, Proposition
\ref{prop:SPtocore}, in conjunction with Theorem~\ref{thm:main}, leads
to viewing the pure form of survey propagation $\SP(1)$ as performing
an approximate marginalization over cores.  Therefore, our results
raise the crucial question: do cores exist for random formulas?
Motivated by this perspective, Achlioptas and Ricci-Tersenghi
\cite{AR05} has answered this question affirmatively for $k$-SAT with
$k\ge9$.  In Section~\ref{SecRandEns}, we show that cores, if they
exist, must be ``large'' in a suitable sense (see
Proposition~\ref{PropLargeCores}).  In the following section, we
explore the case $k=3$ via experiments on large instances.

\vtiny

\subsection{Peeling experiments} 

\begin{figure}
\begin{center}
\widgraph{0.5\textwidth}{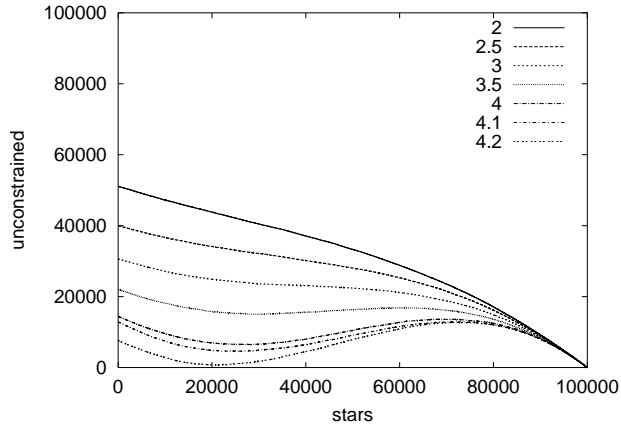} 
\caption{Evolution of the number of unconstrained variables in the
peeling process: start with a satisfying assignment, change a random
unconstrained variable to $*$ and repeat. Plotted is the result of an 
experiment for $n=100,000$, for random formulas with $k=3$ and
$\alpha=\{2, 2.5, 3, 3.5, 4, 4.1, 4.2\}$.  In particular, core
assignments are on the $x$-axis, and satisfying assignments are on the
$y$-axis.}
\label{fig:peel}
\end{center}
\end{figure}

We have performed a large number of the following experiments:
\begin{enumerate}
\item starting with a satisfying assignment $x$,
change a random one of its unconstrained variables to $*$, 
\item repeat until there are no unconstrained variables.
\end{enumerate}
This procedure, which we refer to as ``peeling'', is equivalent to
taking a random path from $x$ in $\graph$, by choosing at each step a
random outgoing edge. Any such path terminates at the core
$\Core{V}{x}$. It is interesting to examine at each step of this
process the number of unconstrained variables (equivalently, the
number of outgoing edges in the graph $\dirgraph$).  For $k=3$ SAT
problems, Figure~\ref{fig:peel} shows the results
of such experiments for $n=100,000$, and using
different values of $\alpha$. The plotted curves are the evolution of
the number of unconstrained variables as the number of $*$'s
increases.
We note that for $n=100$ and $\alpha$ close to the threshold, satisfying
assignments often correspond to core assignments; a similar
observation was also made by Braunstein and Zecchina~\cite{BZ04}.  In
contrast, for larger $n$, this correspondence is rarely the case.
Rather, the generated curves suggest that $\Core{V}{x}$ is almost
always the all-$*$ assignment, and moreover that for high density
$\alpha$, there is a critical level in $\graph$ where the out-degrees
are very low. Increasing $\alpha$ results in failure of the algorithm
itself, rather than in the formation of real core assignments.

For $k=2$, the event that there is a path in $\graph$ from a
satisfying assignment to the all-$*$ assignment has a very natural
interpretation.  In particular, it is equivalent to the event that the
\emph{pure-literal rule} succeeds in finding an assignment. The
pure-literal rule~\cite{RPF99} is an algorithm consisting of the
following steps: assign $1$ to a variable if it only appears
positively in a clause, and $0$ if it only appears negatively in a
clause, reduce the formula, and repeat the procedure.  It is
straightforward to check that the sequence of variables given by the
labels on any path from the all-$*$ assignment to a satisfying
assignment can be identified with a sequence of steps of the
pure-literal type.  Furthermore, it is known~\cite{RPF99} that there
is a phase transition for the event that the pure-literal rule
succeeds at $\alpha=1$.

Interestingly, as mentioned earlier, for $k\ge 9$ there are values for
$\alpha < \alpha_c$ such that this peeling procedure provably results
in a non-trivial core assignment with high probability, according to
\cite{AR05}.  The fact that we do not observe core assignments for
$k=3$, and yet the algorithm is successful, means that an alternative
explanation is required.  Accordingly, we propose studying the
behavior of $\SP(\rho)$ for $\rho \in (0,1)$.  Our experimental
results, consistent with similar reports from
Kirkpatrick~\cite{Kirkpatrick04}, show that $\SP(\rho)$ tends to be
most effective in solving $k$-SAT for values of $\rho < 1$.  If so,
the good behavior of $\SP(1)$ may well follow from the similarity of
$\SP(1)$ updates to $\SP(\rho)$ updates for $\rho \approx 1$.  To
further explore this issue, the effects of varying the weight
distribution $(\wun, \wstar)$, and consequently the parameter $\rho$,
are discussed in the following section.

\vtiny

\subsection{Weight distribution and smoothing} 
\label{subsec:weights}
One of the benefits of our analysis is that it suggests a large pool
of algorithms to be investigated. One option is to vary the values of
$\wun$ and $\wstar$.  A ``good'' setting of these parameters should
place significant weight on precisely those valid assignments that can
be extended to satisfying assignments.  At the same time, the
parameter setting clearly affects the level of connectivity in the
space of valid assignments. Connectivity most likely affects the
performance of belief propagation, as well as any other algorithm that
we may apply to compute marginals or sample from the distribution.

Figure~\ref{fig:parameters}(a) shows the performance of belief
propagation on the extended MRF for different values of $(\wun,
\wstar)$, and applied to particular random formula with $n=10,000$,
$k=3$ and $\alpha=4.2$. The most successful pairs in this case were
$(0.05, 0.95)$, $(0.05, 0.9)$, $(0.05, 0.85)$, and $(0.05, 0.8)$. For these
settings of the parameters the decimation steps reached a solution, so a
call to WalkSAT was not needed. 
For weights satisfying $\wun+\wstar >1$, the
behavior is very predictable: although the algorithm converges, the
choices that it makes in the decimation steps lead to a
contradiction. Note that there is a sharp transition in algorithm
behavior as the weights cross the line $\wun+\wstar = 1$, which is
representative of the more general behavior. 
\begin{figure}
\begin{center}
\psfrag{\wun}{$\wun$} 
\psfrag{\wstar}{$\wstar$}
\begin{tabular}{c}
\widgraph{0.52\textwidth}{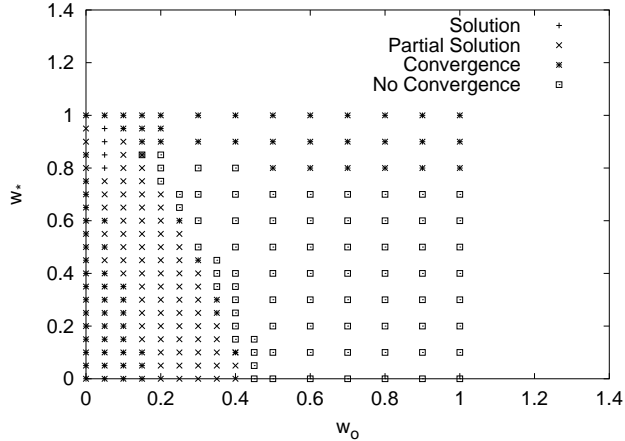} 
\end{tabular}
\caption{Performance of BP for different choices of ($\wun, \wstar)$ as
applied to a particular randomly chosen formula with $n=10000$, $k=3$,
$\alpha=4.2$. Four distinct cases can be distinguished: (i) BP
converges and the decimation steps yields a complete solution, (ii) BP
converges and the decimation steps yield a partial solution, completed
by using Walk-SAT, (iii) BP converges, but the decimation steps don't
lead to a solution, and (iv) BP does not converge.}
\label{fig:parameters}
\end{center}
\end{figure}
\comment{
\begin{figure}
\begin{center}
\psfrag{\wun}{$\wun$}
\psfrag{\wstar}{$\wstar$}
\begin{tabular}{cc}
\widgraph{0.52\textwidth}{par.eps} &
\widgraph{0.45\textwidth}{fig_decimation_Long1.eps} \\
(a) & (b)
\end{tabular}
\caption{(a) Performance of BP for different settings of $\wun$, and
$\wstar$ for a particular random formula with $n=10,000$, $k=3$,
$\alpha=4.2$. We distinguish between four cases: (i) BP converges and
the decimation steps yields a complete solution, (ii) BP converges and
the decimation steps yield a partial solution, completed by using
Walk-SAT, (iii) BP converges but the decimation steps don't lead to a
solution, and (iv) BP does not converge.  (b) Gibbs and decimation: In
this experiment on $n=400$ variables and $\alpha=4.1$ we draw the
trajectory of the dynamics in the $(\mbox{number of } *,\mbox{ number
of unconstrained})$ axis. For comparison we also draw the peeling
curve corresponding to an assignment found by SP.  The dynamics were
run with $10^7$ steps between decimations where at each decimation
step a single variable is set.}
\label{fig:parameters}
\end{center}
\end{figure}
}

The following result provides some justification for the
excellent performance in the regime $\wun + \wstar \leq 1$.
\begin{theorem}
\label{ThmSumOne}
If $\wun+\wstar = 1$, then $\sum_{y \leq x} \Weight(y) =
\wstar^{n_*(x)}$ for any valid assignment $x$.
If $\wun+\wstar < 1$, then $\sum_{y \leq x} \Weight(y) \geq
(\wstar)^{n_*(x)}$ for any valid assignment $x$.
\end{theorem}

It should be noted that Theorem~\ref{ThmSumOne} has a very natural
interpretation in terms of a ``smoothing'' operation.  In particular,
the $(\wun, \wstar)$-MRF may be regarded as a smoothed version of the
uniform distribution over satisfying assignments, in which the uniform
weight assigned to each satisfying assignment is spread over the
lattice associated with it.\footnote{Note, however, that any
partial assignment that belongs to two or more lattices is
assigned a weight only once.  Otherwise, the transformation would be a
convolution operation in a strict sense.}

The remainder of this section is devoted to the proof of Theorem 
\ref{ThmSumOne}.

\begin{proof}
We start with the case $\wun+\wstar = 1$.
Let $\Consistent{x}$ denote the set of partial assignments $z$
such that $z_j \in \{x_j, *\}$ for all $j \in \vertexset$. We refer to
these as the set of assignments consistent with $x$. Let
$\Reachable{x}=\{y: y\le x\}$ be the set of valid assignments that are
reachable from $x$. Notice that all $y\in \Reachable{x}$ are valid and
consistent with $x$, but not every valid assignment in
$\Consistent{x}$ is reachable from $x$. We will let $\Setstar(z)$ denote the
set of variables assigned $*$ both for valid and invalid assignments
$z$.

We define a map between all assignments consistent with $x$ and the
set of reachable ones.  Let $\MaptoReachable: \Consistent{x} \ra
\Reachable{x}$ be defined as
\begin{equation*}
\MaptoReachable(z) \defn  \Core{{\Setstar(z)}}{x}.
\end{equation*}
Notice that if $y \in \Reachable{x}$ then $\MaptoReachable(y)=y$. 
The map is, of course, many-to-one. 
We define what we'll show is the reverse map. For $y\in \Reachable{x}$ let 
\begin{equation*}
\MapfromReachable(y) \defn \{z \in \Consistent{x} :
\Setstar(z) = \Setstar(y) \cup T , T \subseteq \Setcon(y)\}.
\end{equation*}

\begin{lemma}
\label{lemma:bijection}
For any $y\in \Reachable{x}$ and $z\in \Consistent{x}$, 
$z\in \MapfromReachable(y)$ if and only if $\MaptoReachable(z)=y$.
\end{lemma}
\begin{proof}
Let $z \in \MapfromReachable(y)$ so that $\Setstar(z)=\Setstar(y) \cup
T$ for some $T \subseteq \Setcon(y)$. 
$\MaptoReachable(z)=\Core{{\Setstar(z)}}{x}$ is the
minimal valid assignment such that the path from $x$ to it is labeled
only by elements in $\Setstar(z)$. We'll show that $y$ satisfies these
properties, and therefore by proposition~\ref{prop:core},
$y=\MaptoReachable(z)$.  Any path from $x$ to $y$ (which exists since
$y\in \Reachable{x}$) is labeled by $\Setstar(y)\backslash \Setstar(x)
\subseteq \Setstar(z)$. Furthermore, for every $i \in \Setstar(z)$, $i\notin
\Setunc(y)$ so there is no outgoing edge from $y$ labeled by an element in
$\Setstar(z)$. Therefore $y$ is minimal.

Let $y=\MaptoReachable(z)=\Core{{\Setstar(z)}}{x}$. By
proposition~\ref{prop:core} there is no $i \in \Setstar(z)$ such that
$i\in \Setunc(y)$. Therefore 
$\Setstar(z) \subseteq \Setstar(y) \cup \Setcon(y)$. 
Further we have that 
$\Setstar(y) \subseteq \Setstar(z) \cup \Setstar(x)=\Setstar(z)$, 
therefore $\Setstar(z)=\Setstar(y)\cup T$ for some
$T \subseteq \Setcon(y)$. Hence $z\in \MapfromReachable(y)$.
\end{proof}

\begin{figure}
\label{fig:bijection}
\begin{center}
\scalebox{0.4}{\includegraphics{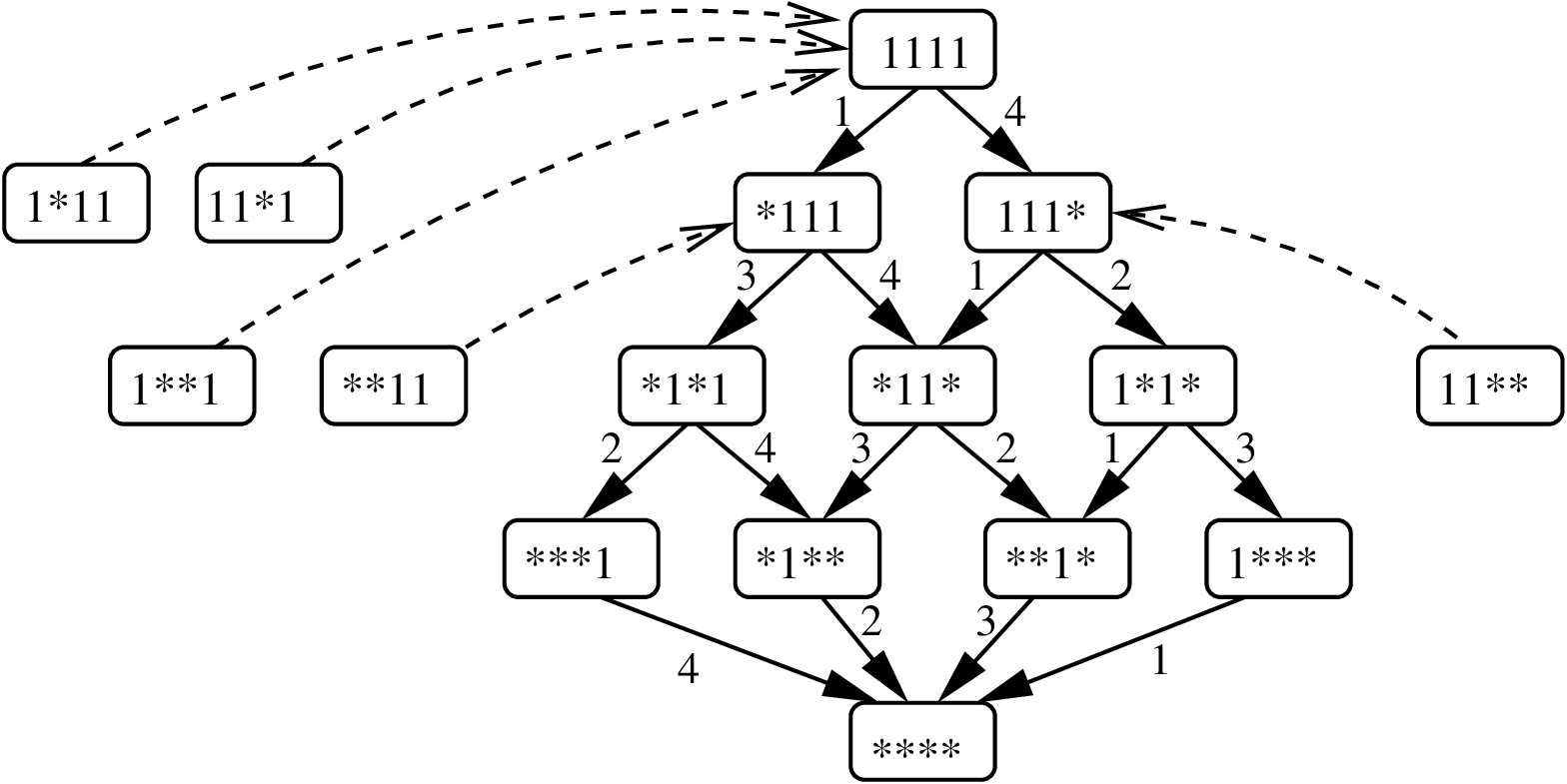}}
\caption{The directed graph $\graph$ and the map $\MaptoReachable$ for the
formula 
$(x_1 \vee x_2\vee x_3)\wedge ( \bar{x_2}\vee \bar{x_3}\vee x_4)$ 
and the satisfying assignment $(0,0,1,0)$. The solid arrows denote
edges in $\graph$ and the dashed arrows denote $\MaptoReachable$.}
\end{center}
\end{figure}

For a set of partial assignments $X$ let $\Weight(X)=\sum_{x \in
X}\Weight(x)$.  Let $\Weight^\emptyset (z) = (\wstar)^{n_*(z)} \times
(\wun)^{n-n_*(z)}$, denote the weight of any partial assignment,
if the formula had no clauses. For such a formula all partial
assignments are valid.  Observe that if we restrict our attention to
the assignments that are consistent with $x$,
\begin{eqnarray*}
\Weight^\emptyset (\Consistent{x}) 
&=& \sum_{z \in \Consistent{x}} \Weight^\emptyset(z) \\
&=& \sum_{S \subseteq V \backslash \Setstar(x)} (\wstar)^{|\Setstar(x)|+|S|} 
\times (\wun)^{n - |\Setstar(x)| - |S|} \\ 
&=& (\wstar)^{|\Setstar(x)|} \times (\wstar+\wun)^{n-|\Setstar(x)|} \\
&=& (\wstar)^{n_*(x)}
\end{eqnarray*}

We show that when clauses are added to the formula, the total weight
under $x$ is preserved as long as $x$ is still valid.  In particular
when an assignment $z$ that is consistent with $x$ becomes invalid, it
passes its weight to an assignment that is still valid, namely
$\MaptoReachable(z)$, which has fewer $*$ variables than $z$.
\begin{eqnarray}
\Weight (y) &=& (\wstar)^{n_*(y)} \times (\wun)^{n_\unc(y)} \times
1^{n_c(y)} \nonumber \\
\label{eq:weights}
&=& (\wstar)^{n_*(y)} \times (\wun)^{n_\unc(y)} \times (\wstar +
\wun)^{n_c(y)} \\
&=& \sum_{T \subseteq \Setcon(y)} (\wstar)^{n_*(y) +|T|} \times
(\wun)^{n_\unc(y) + n_c(y) - |T|}  \nonumber \\
&=& \sum_{T \subseteq \Setcon(y)} \Weight^\emptyset( z :
\Setstar(z)=\Setstar(y) \cup T ) \nonumber \\ 
&=& \Weight^\emptyset( \{z : \Setstar(z)=\Setstar(y) \cup T, T
\subseteq \Setcon(y)\} ) \nonumber \nonumber \\
&=& \Weight^\emptyset(\MapfromReachable(y)). \nonumber
\end{eqnarray}

Finally, we have:
\begin{equation*}
\sum_{y \leq x}\Weight(y) = \sum_{y \leq
x}\Weight^\emptyset(\MapfromReachable(y)) =
\Weight^\emptyset(\Consistent{x}) = (\wstar)^{n_*(x)}
\end{equation*}
where we used the fact that the sets $\MapfromReachable(y)$ for $y\in
\Reachable{x}$ partition $\Consistent{x}$ by lemma~\ref{lemma:bijection}.

The proof of the case $\wun+\wstar < 1$ is similar except that equation
\eqref{eq:weights} becomes an inequality:
\begin{equation*}
\Weight (y)= (\wun)^{n_\unc(y)} \times (\wstar)^{n_*(y)} \times
1^{n_c(y)} \geq \sum_{T \subseteq \Setcon(S)}
\Weight^\emptyset(\MapfromReachable(y)).
\end{equation*}
When an assignment $z$ that is consistent with $x$ becomes invalid, it
passes more than its own weight to $\MaptoReachable(z)$.
\end{proof}

\vtiny

\subsection{Gibbs sampling} 
\label{SecGibbs}

Based on our experiments, the algorithm $\SP(\rho)$ is very effective
for appropriate choices of the parameter $\rho$. The link provided by
Theorem~\ref{ThmSumOne} suggests that the distribution $p_{\Weight}$,
for which $\SP(\rho)$---as an instantiation of belief propagation on
the extended MRF---is computing approximate marginals, must posses
good ``smoothness'' properties.  One expected consequence of such
``smoothness'' is that algorithms other than BP should also be
effective in computing approximate marginals.  Interestingly, rigorous
conditions that imply (rapid) convergence of BP~\cite{TJ02}---namely,
uniqueness of Gibbs measures on the computation tree---are quite
similar to conditions implying rapid convergence of Gibbs samplers,
which are often expressed in terms of ``uniqueness'', ``strong spatial
mixing'', and ``extremality'' (see, for
example~\cite{Martinelli99,BKMP04}).

In this section, we explore the application of sampling methods to the
extended MRF as a means of computing unbiased stochastic
approximations to the marginal distributions, and hence biases at each
variable.
\begin{figure}[h]
\begin{center}
{\small{
\begin{tabular}{cc}
\begin{tabular}{|c||c|c|c|c|}
\hline 
\multicolumn{1}{|c||}{SAT $\alpha$}  & \multicolumn{4}{|c|}{Gibbs $\rho$} \\
\hline
& $\quad 0.4 \quad$ & $\quad 0.5 \quad$ & $\quad 0.7 \quad$ & $\quad 0.9
\quad$ \\ %
\hline \cline{2-5}
4.2      &    {\bf{0.0493}}     &   0.1401    &  0.3143     & 0.4255 \\ \hline
4.1      &    {\bf{0.0297}}     &   0.1142   &   0.3015     & 0.4046 \\ \hline
4.0      &    0.0874     &   {\bf{0.0416}}   &   0.2765     & 0.3873 \\ \hline
3.8      &    0.4230     &   0.4554   &   0.1767     & {\bf{0.0737}} \\ \hline
3.6      &    0.4032     &   0.4149   &   0.1993     & {\bf{0.0582}} \\ \hline
3.4      &    0.4090     &   0.4010   &   0.2234     & {\bf{0.0821}} \\ \hline \hline
\end{tabular}
&
\begin{tabular}{|c||c|c|c|c|}
\hline 
\multicolumn{1}{|c||}{SAT $\alpha$}  & \multicolumn{4}{|c|}{Gibbs $\rho$} \\
\hline
& $\quad 0.4 \quad$ & $\quad 0.5 \quad$ & $\quad 0.7 \quad$ & $\quad 0.9
\quad$ \\ %
\hline \cline{2-5}

4.2  &          {\bf{0.0440}}  &  0.1462 &     0.3166 &    0.4304 \\ \hline 
4.1  &          0.0632 &  {\bf{0.0373}} &     0.2896 &    0.4119 \\ \hline 
4.0  &          {\bf{0.0404}} &  0.0666 &     0.2755 &    0.3984 \\ \hline 
3.8  &          0.1073 &  {\bf{0.0651}} &     0.2172 &    0.3576 \\ \hline 
3.6  &          0.1014 &  {\bf{0.0922}} &     0.1620 &    0.3087 \\ \hline 
3.4  &          0.3716 &  0.3629 &     0.1948 &    {\bf{0.0220}} \\ \hline \hline
\end{tabular} \\
& \\
(a) Comparison to $\SP(0.95)$ & (b) Comparison to $\SP(0.9)$
\end{tabular}

\vtiny

\begin{tabular}{cc}
 & \\
\begin{tabular}{|c||c|c|c|c|}
\hline 
\multicolumn{1}{|c||}{SAT $\alpha$}  & \multicolumn{4}{|c|}{Gibbs $\rho$} \\
\hline
& $\quad 0.4 \quad$ & $\quad 0.5 \quad$ & $\quad 0.7 \quad$ & $\quad 0.9
\quad$ \\ %
\hline \cline{2-5}

4.2 &         SP fails   &  SP fails &   SP fails  & SP fails \\ \hline
4.1 &         {\bf{0.0230}}   &     0.0985  &    0.3236   &  0.4341 \\ \hline
4.0 &         0.0493   &     {\bf{0.0079}}   &   0.3273   &  0.4309 \\ \hline
3.8 &         0.0531   &     {\bf{0.0194}}   &   0.2860   &  0.4104 \\ \hline
3.6 &         0.0980   &     {\bf{0.0445}}   &   0.2412   &  0.3887 \\ \hline
3.4 &         0.0365   &     {\bf{0.0356}}  &    0.1301   &  0.3869 \\ \hline \hline
\end{tabular} 
& 
\begin{tabular}{|c||c|c|c|c|}
\hline
\multicolumn{1}{|c||}{SAT $\alpha$}  & \multicolumn{4}{|c|}{Gibbs $\rho$} \\
\hline
& $\quad 0.4 \quad$ & $\quad 0.5 \quad$ & $\quad 0.7 \quad$ & $\quad 0.9
\quad$ \\ %
\hline \cline{2-5}

4.2  &           SP fails    & SP fails      &  SP fails    &  SP fails \\ \hline
4.1 &            {\bf{0.1925}}  &  0.2873   &   0.3989  &   0.4665 \\ \hline
4.0 &            {\bf{0.0483}}  &  0.1092   &   0.2986  &   0.4179 \\ \hline
3.8 &            0.0924  &  {\bf{0.0372}}   &   0.3235  &   0.4323 \\ \hline
3.6 &            {\bf{0.0184}}  &  0.0304   &   0.2192  &   0.4009 \\ \hline
3.4 &            0.0323  &  {\bf{0.0255}}   &   0.0718  &   0.3613  \\ \hline \hline
\end{tabular}  \\
&  \\

(c) Comparison to $\SP(0.7)$
& (d) Comparison to $\SP(0.5)$
\end{tabular}
}}
\caption{Comparison of $SP(\beta)$ pseudomarginals for $\beta \in
\{0.95, 0.9, 0.7, 0.5 \}$ to marginals estimated by Gibbs sampling on
weighted MRFs with $\rho \in \{0.4, 0.5, 0.7, 0.9 \}$ for the range of
SAT problems $\alpha \in \{4.2, 4.1, 4.0. 3.8, 3.6, 3.4\}$.  Each
entry in each table shows the average $\ell_1$ error between the
biases computed from the $SP(\beta)$ pseudomarginals compared to the
biases computed from Gibbs sampling applied to $MRF(\rho)$.
Calculations were based on top 50 most biased nodes on a problem of
size $n = 1000$. The bold entry within each row (corresponding to a
fixed $\alpha$) indicates the $MRF(\rho)$ that yields the smallest
$\ell_1$ error in comparison to the SP biases.}
\label{FigGibbsMarg}
\end{center} 
\end{figure}
More specifically, we implemented a Gibbs sampler for the family of
extended MRFs developed in Section~\ref{sec:MRF}.  The Gibbs sampler
performs a random walk over the configuration space of the extended
MRF---that is, on the space of partial valid assignments.  Each
step of the random walk entails picking a variable $x_i$ uniformly at
random, and updating it randomly to a new value $b \in \{0,1,*\}$
according to the conditional probability $p_{\Weight}(x_i = b | (x_j :
j \neq i))$.  By the construction of our extended MRF (see
equation~\eqref{EqnDefnSuperMRF}), this conditional probability is an
(explicit) function of the variables $x_j$ and $x_i$ appear together
in a clause, and of the variables $x_k$ such that $x_k$ and $x_j$
appear together in a clause, where $x_j$ and $x_i$ appear together in
a clause.

It is of interest to compare the approximate marginals computed by the
$\SP(\beta)$ family of algorithms (to which we refer as
{\emph{pseudomarginals}}) to the (stochastic) estimates computed by
Gibbs sampler.  Given the manner in which the SP pseudomarginals are
used in the decimation procedure, the most natural comparison is
between the biases $\spmarg_i(0)- \spmarg_i(1)$ provided by the
$SP(\beta)$ algorithm, and the biases $\tau_i(0) - \tau_i(1)$
associated with the Gibbs sampler (where $\tau_i$ are the approximate
marginals obtained from Gibbs sampling on the extended MRF with
parameter $\rho$ (denoted $\MRF(\rho)$).  The results of such
comparisons for the SP parameter $\beta \in \{0.95, 0.9, 0.7, 0.5 \}$
and the Gibbs sampling parameter $\rho \in \{0.4, 0.5, 0.7, 0.9 \}$
are shown in Figure~\ref{FigGibbsMarg}.  Comparisons are made for each
pair $(\beta, \rho)$ in these sets, and over a range of clause
densities $\alpha \in \{4.2, 4.1, 4.0. 3.8, 3.6, 3.4\}$.  For fairly
dense formulas (e.g., $\alpha \geq 4.0$), the general trend is that
the $\SP(\beta)$ biases with larger $\beta$ agree most closely with
the Gibbs biases with $\rho$ relatively smaller (i.e., $\rho <
\beta$).  For lower clause densities (e.g., $\alpha = 3.4$), the
agreement between the $\SP(\beta)$ and Gibbs biases on $MRF(\rho)$
when $\beta = \rho$ is substantially closer.  


\comment{ MJW --- this text and the associated figure are perhaps
tangential now.  Our experiments with Gibbs sampling are still at
their initial stage.  We have succeeded to solve a number of random
$3$-SAT problems using Gibbs sampling and decimation for $n=1000$ with
$\alpha=4$.
See Figure~\ref{fig:parameters}(b) for a typical run of Gibbs sampling
and decimation.  Note how the points sample during the course of Gibbs
sampling and decimation follow the peeling curve back towards the
$y$-axis, corresponding to a satisfying assignment.  This behavior
provides further support for the significance of the extended MRF
representation in the performance of survey-based decimation.
}


\vtiny 

\section{Expansion arguments for random formulas} 
\label{SecRandEns}

\newcommand{\Elprob}{\ensuremath{\mathbb{P}}}
\newcommand{\Elexp}{\ensuremath{\mathbb{E}}}

This section is devoted to the study of properties of the MRF on
random formulas. We will use simple random graph arguments in order to
obtain typical properties of cores, as well as the behavior of Gibbs
sampling or message-passing algorithms applied to the MRF associated
with a randomly chosen formula.  Throughout this section, we denote
$p^{\phi}_{\Weight}$ to denote the MRF distribution for a fixed
formula $\phi$.  Otherwise, we write $\Elprob^{n,m}$ for the uniform
measure on $k$-sat formulas with $n$ variables and $m$ clauses, and
$\Elprob^{n, \alpha}$ for the uniform measure on $k$-sat formulas with
$n$ variables and $m = \alpha n$ clauses.  We often drop $n$, $m$,
and/or $\alpha$ when they are clear from the context.  Finally, we use
$\E^{\phi}_{\Weight}$, $\Elexp^{n,m}$ and $\Elexp^{n, \alpha}$ to
denote expectations with respect to the distributions
$p^{\phi}_{\Weight}$, $\Elprob^{n,m}$ and $\Elprob^{n, \alpha}$
respectively.

\vtiny
\subsection{Size of cores}

We first prove a result that establishes that cores, if they exist,
are typically at least a certain linear fraction $\crat(\alpha,k)$ of
the total number $n$ of variables.
\begin{proposition} 
\label{PropLargeCores}
Let $\Kform$ be a random $k$-sat formula with $m=\alpha n$ clauses
where $k \geq 3$.  Then for all positive integers $C$ it holds that 
\begin{eqnarray}
\Elprob^{n, \alpha} [\mbox{ $\Kform$ has a core with $\csize$ clauses
 }] & \leq & \left( \frac{e^2 \alpha \csize^{k-2}}{n^{k-2}}
 \right)^\csize,
\end{eqnarray}
Consequently, if we define $\crat(\alpha, k) \defn (\alpha
e^2)^{-1/(k-2)}$, then with $\Elprob^{n, \alpha}$-probability tending
to one as $n \rightarrow +\infty$, there are no cores of size strictly
less than $\crat(\alpha, k) \, n$.
\end{proposition}
\begin{proof}
Suppose that the formula $\Kform$ has a core with $\csize$ clauses.
Note that the variables in these clauses all lie in some set of at
most $\csize$ variables.  Thus the probability that a core with
$\csize$ clauses exist is bounded by the probability that there is a
set of $\csize$ clauses all whose variables lie in some set of size
$\leq \csize$.  This probability is bounded by
\[
\binom{m}{\csize} \binom{n}{\csize}
\left(\frac{\csize}{n}\right)^{\csize k},
\]
which can be upper bounded by
\[
\left( \frac{e m}{\csize} \right)^{\csize} \left( \frac{e n}{\csize}
\right)^{\csize} \left( \frac{\csize} {n} \right)^{\csize k} = \left(
\frac{e^2 \alpha \csize^{k-2}}{n^{k-2}} \right)^\csize,
\]
as needed.
\end{proof}

\vtiny

\subsection{(Meta)-stability of the all $*$ assignment for small
  $\rho$}

By definition, the extended MRF for $\rho = 1$ assigns positive mass
to the all-$\ast$ vector. Moreover, Proposition~\ref{PropLargeCores}
implies that the size of cores (when they exist) is typically linear
in $n$. It follows that the state space of the MRF for $\rho = 1$
typically satisfies one of the following properties:
\begin{itemize}
\item
Either the state space is trivial, meaning that it contains only the
all $*$ state, or
\item
The state space is disconnected with respect to all random walks based
on updating a small linear fraction of the coordinates in each step.
\end{itemize}
The goal of this section is to establish that a similar phenomenon
persists when $\rho$ is close to $1$ (i.e., when $1-\rho$ is small).

We begin by introducing some notions from the analysis of the mixing
properties of Markov chains.  Let $T$ be a reversible chain with
respect to a measure $p$ on a state space $\Omega$.  For sets $A, B
\subset \Omega$, write
\[
q_T(A,B) = \sum_{x \in A, y \in B} p(x) T_{x \to y} = \sum_{x \in A, y
           \in B} p(y) T_{y \to x}.
\] 
The \emph{conductance} of the chain $T$ is defined as
\[
c(T) = \inf_{S \subset \Omega} \{ \frac{q_T(S,S^c)}{p(S)(1-p(S))}\}.
\]
It is well-known that $c(T)/2$ is an upper bound on the spectral gap
of the chain $T$ and that $2/c(T)$ is a lower bound on the mixing time
of the chain. We note moreover that the definition of $T$ implies that
for every two sets $A,B$ it holds that $q_T(A,B) \leq
\min\{p(A),p(B)\}$.

\begin{definition} \label{def:bad_mixing}
Consider a probability measure $p$ on a space $\Omega$ of strings of
length $n$. Let $T$ be a Markov chain on $\Omega$. The radius of $T$
denoted by $r(T)$ is defined by
\begin{equation}
r(T) \defn \sup \{d_H(x,y) : T_{x,y} > 0\},
\end{equation}
where $d_H$ is the Hamming distance.  We let the radius
$r$-conductance of $p$ denote by $c(r,p)$ be
\begin{equation}
c(r,p) \defn \sup \{c(T) : \mbox{ T is reversible with respect to } p
  \mbox{ and } r(T) \leq r\}.
\end{equation}
\end{definition}

Now returning to the random $k$-SAT problem, we write $p_{\rho}$ for
the measure $p_{\Weight} = p^\phi_{\Weight}$ with $\wstar = \rho$ and
$\wun = 1-\rho$.
\begin{proposition} \label{prop:bad_mixing}
Consider a randomly chosen $k$-SAT formula with density $\alpha$. Then
there exists a $\rho_0 \in (0,1)$ such that if $\rho > \rho_0$ then
$\Elprob^n[\phi \in A_n \cup B_n] \to 1$ as $n \rightarrow +\infty$
where $A_n$ and $B_n$ are the following events:
\begin{enumerate}
\item[(I)] $A_n$ consists of all the formulas $\phi$ satisfying
$p_\rho^{\phi}[n - n_*(x) \leq 2\sqrt{(1-\rho)} \, n] \geq 1 -
\exp(-\Omega(n))$.
\item[(II)] $B_n$ consists of all the formulas $\phi$ for which the
measure $p_{\rho}^{\phi}$ satisfies $c(\sqrt{(1-\rho)} \, n, p_\rho)
\leq \exp(-\Omega(n))$.
\end{enumerate}
\end{proposition}

\begin{proof} 
We let $\delta$ be a small positive number to be determined, and set
$1-\rho = \delta^2$.  As it suffices to work with ratios of
probabilities, we use the unnormalized weight $W^{\phi}(x)$ instead of
$p^{\phi}_{\Weight}(x)$.

The proof requires the following:
\begin{lemma}
\label{LemCond}
Let $d$ be an integer satisfying $\delta n \leq d \leq 2 \delta n$.
For $\delta$ sufficiently small, it holds that with $\Elprob^n$
probability going to $1$ as $n \to \infty$
\begin{equation} 
\label{eq:cond}
\frac{\sum_{d = \delta n}^{2 \delta n} W^{\phi}[n-n_* = d]}{\rho^{3n}}
  = \exp(-\Omega(n)).
\end{equation}
\begin{proof}
See Appendix~\ref{AppLemCond}.
\end{proof}
\end{lemma}

To establish the proposition, it suffices to show that for any formula
$\phi$ for which equation~\eqref{eq:cond} of Lemma~\ref{LemCond} is
valid, then one of either condition (I) or condition (II) must hold.
\begin{enumerate}
\item[(i)] First suppose that $W^{\phi}[n-n_*(x) > 2 \delta n] \leq
\rho^{3n/2}$. In this case, condition (I) in the statement of the
proposition follows immediately.
\item[(ii)] Otherwise, we may take $W^{\phi}[n-n_*(x) > 2 \delta n]
\geq \rho^{3n/2}$. In this case, we can apply the conductance bound in
order to bound the gap of any operator with radius $\leq \delta
n$. Take the set $A$ to be all $x$ with $n-n_*(x) < \delta n$ and $B$
be the set of all $x$ with $\delta n \leq n-n_*(x) \leq 2 \delta
n$. Let $T$ be any Markov chain with radius $\delta n$ that is
reversible with respect to $p_{\Weight}$.  Then we have $q_T(A,A^c) =
q_T(A,B) \leq p(B)$.  In addition, it holds that $\Weight^\phi[n -
n_*(x) < \delta n] \geq \rho^n$ (since if $x$ is the all-$\ast$
assignment, we have $\Weight^\phi(x) = \rho^n$); moreover, if we take
$n$ sufficiently large, then we have $\Weight^\phi[ \delta n \leq n -
n_*(x) \leq 2 \delta n] \leq \rho^{3n}$ by Lemma~\ref{LemCond}.
Combining these inequalities, we obtain that the conductance of $T$ is
bounded above by
\begin{eqnarray*}
\frac{q(A,A^c)}{p(A)p(A^c)} & \leq & \frac{p(B)}{p(A)p(A^c)} \\
& \leq & \frac{W^{\phi}[\delta n \leq n-n_*(x) \leq 2 \delta n]}
{W^{\phi}[n-n_*(x) < \delta n] W^{\phi}[n-n_*(x) > 2 \delta n]} \\
& \leq & \frac{\rho^{3n}}{ \rho^n \; \; \rho^{\frac{3n}{2}}} \; = \;
 \rho^{n/2},
\end{eqnarray*}
which implies condition (II).
\end{enumerate}
\end{proof}

\vtiny

\subsection{Message-passing algorithms on random ensembles}

The analysis of the preceding section demonstrated that for values of
$\rho$ close to $1$, any random sampling technique based on local
moves (e.g., Gibbs sampling), if started at the all $*$ assignment,
will take exponentially long to get to an assignment with more than a
negligible fraction of non-$*$.  This section is devoted to
establishing an analogous claim for the belief propagation updates on
the extended Markov random fields.  More precisely, we prove that if
$\rho$ is sufficiently close to $1$, then running belief propagation
with initial messages that place most of their mass on on $*$ will
result assignments that also place most of the mass on $*$.

This result is proved in the ``density-evolution''
setting~\cite[e.g.,]{Richardson01a} (i.e., the number of iterations is
taken to be less than the girth of the graph, so that cycles have no
effect).  More formally, we establish the following:
\begin{theorem} \label{thm:BPall*}
For every formula density $\alpha > 0$, arbitrary scalars $\eps'' > 0$
and $\delta > 0$, there exists $\rho' < 1$, $\eps' \in (0, \eps'')$
and $\gamma > 0$ such that for all $\rho \in (\rho', 1]$ and $\eps \in
(0, \eps')$, the algorithm $SP(\rho)$ satisfies the following
condition.  

Consider a random formula $\phi$, a random clause $b$ and a random
variable $i$ that belongs to the clause $b$.  Then with probability at
least $1-\delta$, if $SP(\rho)$ is initialized with all messages
$\messtovar{a}{j}^0 < \eps$, then the inequality $\messtovar{b}{i}^t <
\eps'$ holds for all iterations $t = 0,1, \ldots,\gamma \log n$.
\end{theorem}

The first step of the proof is to compare the SP iterations to simpler
``sum-product'' iterations.
\begin{lemma} \label{lem:sum_product}
For any $\rho \in [0,1]$, the $SP(\rho)$ iterations satisfy the
inequality:
\begin{eqnarray*} 
\label{eq:sum_product_bound}
\messtovar{\acl}{i}^{t+1} & \leq & \prod_{j \in \Nbrcl(\acl)
\backslash \{i\}} \min \left(1, (1-\rho) + \rho \sum_{b\in \Nbrvar(j)
\backslash \{a\}} \messtovar{b}{j}^t\right)
\end{eqnarray*}
\begin{proof}
See Appendix~\ref{AppLemSumProd}.
\end{proof}
\end{lemma}

Since our goal is to bound the messages $\messtovar{\acl}{i}^{t+1}$,
Lemma~\ref{lem:sum_product} allows us to analyze the simpler
message-passing algorithm with updates specified by:
\begin{equation} 
\label{eq:sum_product}
\messtovar{\acl}{i}^{t+1} = \prod_{j \in \Nbrcl(\acl) \backslash
\{i\}} \min \left(1, (1-\rho) + \rho \sum_{b\in \Nbrvar(j) \backslash
\{a\}} \messtovar{b}{j}^t\right).
\end{equation}
  
The next step is to bound the probability of ``short-cycles'' in the
computation tree corresponding to the message-passing updates
specified in equation~\eqref{eq:sum_product}. More formally, given a
formula $\phi$, we define a directed graph $G(\phi)=(V,E)$, in which
the vertex set $V$ consists of messages $\messtovar{a}{i}$.  The edge
set $E$ includes the edge $\messtovar{a}{i} \to \messtovar{b}{j}$
belongs to $E$ if and only if $j \in \Nbrcl(\acl) \backslash \{i\}$
and $b \in \Nbrdis{\acl}{i}$.  In words, the graph $G(\phi)$ includes
an edge between the $\messtovar{a}{i}$ and $\messtovar{b}{j}$ if the
latter is involved in the update of $\messtovar{a}{i}$ specified in
equation~\eqref{eq:sum_product}.

\begin{lemma} \label{lem:cycles}
Let $G(\phi)$ be the random graph generated by choosing a formula
$\phi$ uniformly at random with $\alpha n$ clauses and $n$ variables.
Let $v$ be a vertex of $G(\phi)$ chosen uniformly at random.  For all
clause densities $\alpha > 0$, there exists $\gamma > 0$ such that
with probability $1-o(1)$, the vertex $v$ does not belong to any
directed cycle of length smaller than $\gamma \log n$ in $G(\phi)$.
\end{lemma}   

\begin{proof} 
The proof is based on standard arguments from random graph
theory~\cite[e.g.,]{Janson00}.
\end{proof}

Our analysis of the the recursion~\eqref{eq:sum_product} on the
computation tree in based on an edge exposure technique that generates
a neighborhood of a vertex $v$ in the graph $G(\phi)$ for a random
$\phi$.  More specifically, pick a clause $a$ and a variable $i$ in
$a$ at random.  Now for each variable $j \in \Nbrcl(\acl) \backslash
\{i\}$, expose all clauses $b$ containing $j$ (but not any other of
the variables appearing so far). Then for each such $b$, we look at
all variables $k \in \Nbrcl(b) \backslash \{j\}$, and so on.  We
consider the effect of repeating this exposure procedure over $t =
\gamma \log n$ steps. When the vertex $\messtovar{a}{i}$ does not
belong to cycles shorter than $t$ in $G(\phi)$, such an analysis
yields a bound on $\messtovar{a}{i}^t$.

\newcommand{\Bin}{\ensuremath{\operatorname{Bin}}}

Note that each clause can expose at most $k-1$ variables. Recall that
we generate the formula $\phi$ by choosing each of the $N_c = 2^k
\binom{n}{k}$ clauses with probability $\alpha n / N_c$.  The
distribution of the number of clauses exposed for each variable is
thus dominated by $\Bin(M_c, \alpha n / N_c)$ where $M_c = 2^k
\binom{n}{k-1}$.  An equivalent description of this process is the
following: each vertex $v = \messtovar{a}{i}$ exposes $X_v$ neighbors
$\messtovar{b}{j}$, where the distribution of the collection $\{X_v\}$
is dominated by a collection $\{Y_v\}$ of i.i.d. random variables.
Moreover, the $Y$'s are jointly distributed as the sum of $k-1$ i.i.d.
$\Bin(M_c, \alpha n / N_c)$ variables.

The proof requires the following lemma on branching processes.
\begin{lemma} 
\label{lem:bp}
Consider a branching process where each vertex gives birth to $Y$
children. Assume further that the branching process is stopped after
$m$ levels and let $K > 0$ be given.

The notion of a {\em good} vertex is defined inductively as follows.
All vertices at level $m$ are good. A vertex at level $m-1$ is good if
it has $\ell$ children and $\ell \leq K$.  By induction for $s \geq 2$
we call a vertex at level $m-s$ good if $v$ has $\ell$ children
$v_1,\ldots,v_\ell$ with $\ell \leq K$ and
\begin{enumerate}
\item[(a)] Either all of $v_1,\ldots,v_\ell$ have at most $K$
children, of which all are good; or
\item[(b)] all of $v_1,\ldots,v_\ell$ have at most $K$ children, of
which all but one are good.
\end{enumerate}
Denote by $p(m,K)$ the probability that the root of the branching
process is good. Then
\[
\inf_{0 \leq m < \infty} p(m,K) = 1 - \exp(-\Omega(K)).
\]
\end{lemma}

\begin{proof}
See Appendix~\ref{AppLemBP}.
\end{proof}

We are now equipped to complete the proof of Theorem \ref{thm:BPall*}.
Using Lemma~\ref{lem:cycles}, first choose $\gamma = \gamma(\alpha)$
such that a random vertex in $G(\phi)$ does not belong to cycles
shorter than $\gamma \log n$ with probability $1-o(1)$.  Next use
Lemma~\ref{lem:bp} to choose $K$ such that the probability $\inf_{0
\leq m < \infty} p(m,K)$ that the root of the branching process is
good is at least $1-\delta/2$.

\newcommand{\newfunc}{\ensuremath{\zeta}}
Next we define a pair of functions $\theta$ and $\newfunc$ (each
mapping $R\times R$ to the real line) in the following way:
\[
\theta(\eps,\rho) \defn ((1-\rho) + K \rho \eps), \qquad \qquad
\newfunc(\eps,\rho) \defn \theta\left(\theta(\eps,\rho),\rho \right)
\times \theta\left(\theta(\eps,\rho)^2,\rho \right).
\] 
Setting $\eps' \defn \min(\eps'',\frac{1}{2K^3})$, observe that
$\theta(\eps',1) = K \eps'$ and therefore $\theta^2(\eps',1) \leq
\frac{\eps'}{4}$ and
\[
\newfunc(\eps',1) = \theta(K \eps',1) \theta((K \eps')^2,1) = (K^2
\eps') (K^4 \eps'^2) = K^6 \eps'^3 \leq \frac{\eps'}{4}.
\]
It now follows by continuity that there exists $\rho' < 1$ such that 
for all $1 \geq \rho \geq \rho'$ it holds that  
\begin{equation} \label{eq:eta_prop}
\begin{array}{ll}
\theta^2(\eps',\rho) \leq \frac{\eps'}{2}, & \newfunc(\eps',\rho) \leq
\frac{\eps'}{2}.
\end{array}
\end{equation}

We claim that the statement of the theorem holds with the choices of
$\gamma,\eps'$ and $\rho'$ above.  Indeed, choose a formula $\phi$
with density $\alpha$ at random and let $v = \messtovar{a}{i}$ be a
random vertex of $G(\phi)$. With probability at least $1 - \delta/2$,
the vertex $v$ does not belong to any cycle shorter than $t = \gamma
\log n$.

Since $v$ does not belong to any such cycle, the first $t$ levels of
the computation tree of $v$ may be obtained by the exposure process
defined above. We will then compare the computation tree to an
exposure process where each variable gives birth to exactly $Bin(M_c,
\alpha n / N_c)$ clauses. Since the messages are generated according
to~\eqref{eq:sum_product}, any bound derived on the values of
non-$\ast$ messages for the larger tree implies the same bound for the
real computation tree.

We now claim that if $v$ is a good vertex on that tree, then the
message at $v$ after $t$ iterations---namely,
$\messtovar{a}{i}^t$----is at most $\eps'$.  Since a vertex of the
tree is good with probability $1-\delta/2$, proving this claim will
establish the theorem.  

We prove this claim by induction on $s$, where $m-s$ is the level of
$w$.  For $s=0$, the claim follows immediately from the initialization
of the messages.  For $s=1$, observe that
equation~\eqref{eq:sum_product} implies that if $w=\messtovar{b}{j}$
is good at level $m-1$, then
\[ 
\messtovar{b}{j} \leq \theta^{k-1}(\rho,\eps) \leq
\theta^2(\rho,\eps') \leq \frac{\eps'}{2}.
\]
For the general induction step, assume that $w=\messtovar{b}{j}$ at
level $m-s$ is good and $s \geq 2$. There are two cases to consider:
\begin{itemize}
\item[(i)] $w$ has all its grand children good. In this case we repeat
the argument above twice to obtain $\messtovar{b}{j} \leq \eps'$.
\item[(ii)] Exactly one of $w=\messtovar{b}{j}$ grand children is not
good.  Let $y'=\messtovar{d'}{\ell'}$ denote the grand-child and
$y=\messtovar{d}{\ell}$ denote $y$ parent.  Then by
equation~\eqref{eq:sum_product}:
\[
\messtovar{d}{\ell} \leq (1-\rho) + K \rho \eps' =
\theta(\eps',\rho).
\]
Using~\eqref{lem:sum_product} again yields
\begin{eqnarray*}
\messtovar{d}{\ell} &\leq& ((1-\rho) + K \rho \theta(\eps',\rho))
((1-\rho) + K \rho \theta^2(\eps',\rho))^{k-2} \\ &\leq& ((1-\rho) + K
\rho \theta(\eps',\rho)) ((1-\rho) + K \rho \theta^2(\eps',\rho)) =
\newfunc(\eps',\rho) \leq \eps'/2,
\end{eqnarray*}
which completes the proof.
\end{itemize}

\vtiny

\section{Conclusion}
\label{SecConclusion}

The survey propagation algorithm, recently introduced by \Mezard,
Parisi and Zecchina~\cite{MPZ02} for solving random instances of
$k$-SAT problems, has sparked a great deal of excitement and research
in both the statistical physics and computer science
communities~\cite[e.g.,]{BMZ03,BMWZ03,BZ04,Aurell05,AR05,MMZ05,Parisi02,WM05}.
This paper provides a new interpretation of the survey propagation
algorithm---namely, as an instance of the well-known belief
propagation algorithm but as applied to a novel probability
distribution over the partial satisfiability assignments associated
with a $k$-SAT formula.
The perspective of this paper reveals the combinatorial structure that
underlies survey propagation algorithm, and we established various
results on the form of these structures and the behavior of
message-passing algorithms, both for fixed instances and over random
ensembles.

The current work suggests various questions and open issues for
further research. As we described, associated with any $k$-SAT problem
is a large family of Markov random fields over partial assignments, as
specified by the parameter $\rho$ (or more generally, the parameters
$\wun$ and $\wstar$).  Further analysis of survey propagation and its
generalizations requires a deeper understanding of the following two
questions.  First, for what parameter choices do the marginals of the
associated Markov random field \emph{yield useful information} about
the structure of satisfiability assignments?  Second, for what
parameter choices do efficient message-passing algorithms like belief
propagation \emph{yield accurate approximations} to these marginals?
Our results show that the success of SP-like algorithms depends on a
delicate balance between these two factors.  (For instance, the
marginals of the uniform distribution over SAT assignments clearly
contain useful information, but belief propagation fails to yield good
approximations for sufficiently large clause densities.)  More
generally, these questions fall in a broader collection of issues, all
related to a deeper understanding of satisfiability problems and
especially the relationship between finite satisfiability problems and
their asymptotic analysis.  Given the fundamental role that
satisfiability plays in diverse branches of computer science, further
progress on these issues is of broad interest.


\section{Acknowledgments}
We would like to thank Dimitris Achlioptas, 
Federico Ardila, Andrea Montanari, Mark \Mezard, 
Giorgio Parisi and Alistair Sinclair for helpful discussions.


\appendix


\section{Belief propagation on a generic factor graph}
\label{AppBP}

Given a subset $S \subseteq \{1,2, \ldots, n\}$, we define $x_S \defn
\{x_i \; | \; i \in S \}$.  Consider a probability distribution on $n$
variables $x_1, x_2, \ldots, x_n$, that can be factorized as
\begin{equation}
\label{eq:prob}
p (x_1, x_2, \ldots, x_n) = \frac{1}{Z} \; \prod_{i=1}^n \psi_i(x_i)
\prod_{a\in \clauseset } \psi_a(x_{V(a)}),
\end{equation}
where for each $a \in \clauseset$ the set $V(a)$ is a subset of $\{1,
2,\ldots n \}$; and $\psi_i(x_i)$ and $\psi_a(x_{V(a)})$ are
non-negative real functions, referred to as compatibility functions,
and
\begin{equation}
Z \defn \sum_{x} \Big [\prod_{i=1}^n \psi_i(x_i) \prod_{a\in
\clauseset } \psi_a(x_{V(a)}) \Big]
\end{equation}
is the normalization constant or partition function.  A factor graph
representation of this probability distribution is a bipartite graph
with vertices $\vertexset$ corresponding to the variables, called {\it
variable nodes}, and vertices $\clauseset$ corresponding to the sets
$V(a)$ and called {\it function nodes}. There is an edge between a
variable node $i$ and function node $a$ if and only if $i \in V(a)$.
We write also $a \in C(i)$ if $i \in V(a)$.

Suppose that we wish to compute the marginal probability of a single
variable $i$ for such a distribution, as defined in
equation~\eqref{EqnDefnMarg}.  The belief propagation or sum-product
algorithm~\cite{Frank01} is an efficient algorithm for computing the
marginal probability distribution of each variable, assuming that the
factor graph is acyclic.  The essential idea is to use the
distributive property of the sum and product operations to compute
independent terms for each subtree recursively. These recursions can
be cast as a message-passing algorithm, in which adjacent nodes on the
factor graph exchange intermediate values. Let each node only have
access to its corresponding compatibility function. As soon as a node
has received messages from all neighbors below it, it can send a
message up the tree containing the term in the computation
corresponding to it.  In particular, let the vectors $\Genmess_{i \ra
a}$ denote the message passed by variable node $i$ to function node
$\acl$; similarly, the quantity $\Genmess_{a \ra i}$ denotes the
message that function node $\acl$ passes to variable node $i$.

The messages from function to variables are updated in the following way:
\begin{eqnarray}
\label{EqnBPClausetoVar}
\Genmess_{a \ra i}(x_i) & \propto & \sum_{x_{\Nbrcl(a) \backslash
\{i\}}} \Big [\compat_a(x_{\Nbrcl(a)}) \prod_{j \in \Nbrcl(a)
\backslash \{i\}} \Genmess_{j \ra a}(x_j) \Big].
\end{eqnarray}
In the other direction, the messages from variable nodes to function
nodes are updated as follows
\begin{eqnarray}
\label{EqnBPVartoClause}
\Genmess_{i \ra a}(x_i) & \propto & \compat_i(x_i) \; \prod_{b \in
\Nbrvar(i) \backslash \{a\} } \Genmess_{b \ra i}(x_i).
\end{eqnarray}
It is straightforward to show that for a factor graph without cycles,
these updates will converge after a finite number of iterations.  Upon
convergence, the local marginal distributions at variable nodes and
function nodes can be computed, using the message fixed point
$\hat{\Genmess}$, as follows:
\begin{subequations}
\begin{eqnarray}
\Field_i(x_i) & \propto & \compat_i(x_i) \; \prod_{b \in \Nbrvar(i)}
\hat{\Genmess}_{b \ra i}(x_i) \\
\Field_a(x_{\Nbrcl(a)}) & \propto & \compat_a(x_{\Nbrcl(a)}) \;
\prod_{j \in \Nbrcl(a)} \hat{\Genmess}_{j \ra a}(x_j).
\end{eqnarray}
\end{subequations}

The same updates, when applied to a graph with cycles, are no longer
exact due to presence of cycles.  An exact algorithm will generally
require exponential time. For certain problems, including
error-control coding, applying belief propagation to a graph with
cycles gives excellent results. Since there are no leaves on graphs
with cycles, usually the algorithm is initialized by sending random
messages on all edges, and is run until the messages converge to some
fixed value~\cite{Frank01}.

\section{Derivation of BP updates on the extended MRF}
\label{AppExtBP}

\subsection{Messages from variables to clauses}

We first focus on the update of messages from variables to clauses.
Recall that we use the notation $\Par{i} = S \cup \{a\}$ as a
shorthand for the event
\begin{eqnarray*}
\acl \in \Par{i} & \mbox{ and } & S = \Par{i}\backslash \{\acl\}
\subseteq \Nbragree{\acl}{i},
\end{eqnarray*}
where it is understood that $S$ could be empty.

\begin{lemma}[Variable to clause messages]
\label{LemVartoClause}
The variable to clause message vector $\Toclmess_{i \ra \acl}$ is
fully specified by values for pairs $(x_i, \Par{i})$ of the form:
\[
\{ 
(\sat{\acl}{i}, S \cup \{a\}), \; \; 
(\sat{\acl}{i}, \emptyset \neq \Par{i} \subseteq \Nbragree{\acl}{i}), \; \; 
(\unsat{\acl}{i}, \emptyset \neq \Par{i} \subseteq \Nbrdis{\acl}{i}), \; \; 
(\sat{\acl}{i}, \emptyset), \; \; 
(\unsat{\acl}{i}, \emptyset), \; \; 
(*, \emptyset) \}.
\]
Specifically, the updates for these five pairs take the following
form:
\begin{subequations}
\label{EqnToclmess}
\begin{eqnarray}
\label{EqnToclmessSatunique}
\Genmess_{i \ra a} (\sat{\acl}{i}, \; \Par{i} = S \cup \{a\} ) & = &
\prod_{b \in S} \tovarsat{b}{i} \; \prod_{b \in \Nbragree{a}{i}
\backslash S} \tovarstar{b}{i} \; \prod_{b \in \Nbrdis{a}{i} }
\tovarun{b}{i} \\
\label{EqnToclmessSatnota}
\Genmess_{i \ra a} (\sat{\acl}{i}, \; \emptyset \neq \Par{i} \subseteq
 \Nbragree{\acl}{i}) & = & 
\prod_{b \in \Par{i}} \tovarsat{b}{i}
\prod_{b \in \Nbragree{\acl}{i} \backslash \Par{i}} \tovarstar{b}{i}\; 
\prod_{b \in \Nbrdis{a}{i} } \tovarun{b}{i} \\
\label{EqnToclmessUnsat}
\Genmess_{i \ra a}(\unsat{\acl}{i}, \; \emptyset \neq \Par{i} \subseteq
\Nbrdis{\acl}{i}) & = & 
\prod_{b \in \Par{i}} \tovarsat{b}{i} 
\prod_{b \in \Nbrdis{a}{i} \backslash \Par{i}} \tovarstar{b}{i} \; 
\prod_{b \in \Nbragree{a}{i}} \; \tovarun{b}{i} \\
\label{EqnToclmessSatempty}
\Genmess_{i \ra a} (\sat{\acl}{i}, \; \Par{i} = \emptyset) & = & \wun
\; \prod_{b \in \Nbragree{a}{i}} \tovarstar{b}{i} 
\; \prod_{b \in \Nbrdis{a}{i}} \tovarun{b}{i} \\
\label{EqnToclmessUnsatempty}
\Genmess_{i \ra a} (\unsat{\acl}{i}, \; \Par{i} = \emptyset) & = & \wun
\; \prod_{b \in \Nbrdis{a}{i}} \tovarstar{b}{i} 
\; \prod_{b \in \Nbragree{a}{i}} \tovarun{b}{i} \\
\label{EqnToclmessStar}
\Genmess_{i \ra a}(*, \Par{i} = \emptyset) & = & \wstar \; \prod_{b
\in \Nbrvar(i) \backslash \{a\}} \tovarstar{b}{i}.
\end{eqnarray}
\end{subequations}
\begin{proof}
The form of these updates follows immediately from the
definition~\eqref{EqnDefnNodeSupcompat} of the variable compatibilities in
the extended MRF, and the BP message update~\eqref{EqnBPVartoClause}.
\end{proof}
\end{lemma}

\subsection{Forms of $R$ quantities}
\label{AppRform}

In this section, we compute the specific forms of the linear sums of
messages defined in equation~\eqref{EqnDefnR}.  First, we use the
definition~\eqref{EqnDefnRsat} and Lemma~\ref{LemVartoClause} to
compute the form of $\Rsat{i}{\acl}$:
\begin{eqnarray*}
\Rsat{i}{\acl} & \defn & \sum_{S \subseteq \Nbragree{\acl}{i}}
\Genmess_{i \ra \acl}(\sat{a}{i}, \Par{i} = S \cup \{a\}) \\
& = & \sum_{S \subseteq \Nbragree{\acl}{i} } \; \; \; \prod_{b \in S}
\tovarsat{b}{i} \; \prod_{b \in \Nbragree{\acl}{i} \backslash S}
\tovarstar{b}{i} \;  \prod_{b \in \Nbrdis{\acl}{i} } \tovarun{b}{i} \\
& = & \prod_{b \in \Nbrdis{a}{i}} \tovarun{b}{i} \Bigl[ \prod_{b \in
\Nbragree{a}{i}} (\tovarsat{b}{i} + \tovarstar{b}{i} ) \Bigr].
\end{eqnarray*}

Similarly, the definition~\eqref{EqnDefnRun} and
Lemma~\ref{LemVartoClause} allows us compute the following form of
$\Run{i}{\acl}$:
\begin{eqnarray*}
\Run{i}{a} & = & \sum_{S \subseteq \Nbrdis{\acl}{i}} \Genmess_{i \ra
a} (\unsat{\acl}{i}, \Par{i}=S) \\
& = & \sum_{S \subseteq \Nbrdis{\acl}{i}, S \neq \emptyset} 
\prod_{b \in S} \tovarsat{b}{i} \; \; 
\prod_{b \in \Nbrdis{\acl}{i} \backslash S} \tovarstar{b}{i} \; \; 
\prod_{b \in \Nbragree{\acl}{i}} \tovarun{b}{i}
+ \wun 
\prod_{b \in \Nbrdis{\acl}{i}} \tovarstar{b}{i} \; \; 
\prod_{b \in \Nbragree{\acl}{i}} \tovarun{b}{i} \\
& = & \prod_{b\in \Nbragree{a}{i}} \tovarun{b}{i} \Bigl[ \prod_{b \in
\Nbrdis{a}{i}} (\tovarsat{b}{i} + \tovarstar{b}{i}) - (1-\wun)
\prod_{b \in \Nbrdis{a}{i}} \tovarstar{b}{i} \Bigr].
\end{eqnarray*}

Finally, we compute $\Rstar{i}{\acl}$ using the
definition~\eqref{EqnDefnRstar} and Lemma~\ref{LemVartoClause}:
\begin{eqnarray*}
\Rstar{i}{a} & = & \Big [\sum_{S \subseteq \Nbragree{a}{i}}
\Genmess_{i \ra a} (\sat{a}{i} , \Par{i} = S) \Big] + \Genmess_{i \ra
a} (*, \Par{i} =\emptyset) \\
& = & \Big[ \sum_{S \subseteq \Nbragree{a}{i}, S \neq \emptyset} \; \;
\prod_{b \in S} \tovarsat{b}{i} \; 
\prod_{b \in \Nbragree{\acl}{i} \backslash S} \tovarstar{b}{i} \; 
\prod_{b \in \Nbrdis{\acl}{i}} \tovarun{b}{i} \Big] 
+ \wun 
\prod_{b \in \Nbragree{\acl}{i}} \tovarstar{b}{i}
\prod_{b \in \Nbrdis{\acl}(i)} \tovarun{b}{i}\\
& & \qquad \qquad \qquad \qquad \qquad \qquad \qquad \qquad + \;\wstar
\prod_{b \in \Nbragree{\acl}{i}} \tovarstar{b}{i} \; 
\prod_{b \in \Nbrdis{\acl}{i}} \tovarstar{b}{i} \\
&=& \prod_{b\in \Nbrdis{\acl}{i}} \tovarun{b}{i} \Bigl[ \prod_{b \in
\Nbragree{\acl}{i}} (\tovarsat{b}{i} + \tovarstar{b}{i}) - (1-\wun)
\prod_{b \in \Nbragree{\acl}{i}} \tovarstar{b}{i} \Bigr] + \wstar
\prod_{b \in \Nbragree{\acl}{i} \cup \Nbrdis{\acl}{i}}
\tovarstar{b}{i}.
\end{eqnarray*}


\subsection{Clause to variable updates}

In this section, we derive the form of the clause to variable updates.
\begin{lemma}[Clause to variable messages]
\label{LemClausetoVar}
The updates of messages from clauses to variables in the extended MRF
take the following form:
\begin{subequations}
\label{EqnClausetoVar}
\begin{eqnarray}
\label{EqnClausetoVarsat}
\tovarsat{a}{i} & = & \prod_{j\in \Nbrcl(a) \backslash \{i\}}
\Run{j}{a} \\
\label{EqnClausetoVarun}
\tovarun{a}{i} & = & \prod_{j \in \Nbrcl(a) \backslash \{i\}}
(\Run{j}{a} + \Rstar{j}{a})+ \sum_{k \in \Nbrcl(a) \backslash \{i\}}
(\Rsat{k}{a} - \Rstar{k}{a} ) \prod_{j\in \Nbrcl(a) \backslash \{i,
k\}} \Run{j}{a} - \prod_{j\in \Nbrcl(a) \backslash \{i\}} \Run{j}{a}
\qquad \quad \\
\label{EqnClausetoVarstar}
\tovarstar{a}{i} & = & \prod_{j\in \Nbrcl(a) \backslash \{i\}}
(\Run{j}{a} + \Rstar{j}{a})- \prod_{j\in \Nbrcl(a) \backslash \{i\}}
\Run{j}{a}.
\end{eqnarray}
\end{subequations}
\end{lemma}
\begin{proof}
(i) We begin by proving equation~\eqref{EqnClausetoVarsat}.  When $x_i
= \sat{a}{i}$ and $\Par{i}=S \cup \{a\}$ for some $S\subseteq
\Nbragree{a}{i}$, then the only possible assignment for the other
variables at nodes in $\Nbrcl(a) \backslash \{i\}$ is $x_j =
\unsat{a}{j}$ and $\Par{j} \subseteq \Nbrdis{a}{j}$.  Accordingly,
using the BP update equation~\eqref{EqnBPClausetoVar}, we obtain the
following update for $\tovarsat{a}{i} \; = \; \Genmess_{a\ra i}
(\sat{a}{i}, \Par{i}=S\cup \{a\})$:
\begin{eqnarray*}
\tovarsat{a}{i} & = & \prod_{j \in \Nbrcl(a) \backslash \{i\}} \; \;
\sum_{\Par{j} \subseteq \Nbrdis{a}{j}} \Genmess_{j \ra a}
(\unsat{a}{j}, \Par{j}) \\ 
& = & \prod_{j\in \Nbrcl(a) \backslash \{i\}} \Run {j}{a}.
\end{eqnarray*}

\vtiny

(ii) Next we prove equation~\eqref{EqnClausetoVarstar}.  In the case
$x_i = *$ and $\Par{i} = \emptyset$, the only restriction on the other
variables $\{ x_j : \; j \in \Nbrcl(a) \backslash \{i\} \}$ is that
they are not all unsatisfying.  The weight assigned to the event that
they are all unsatisfying is
\begin{eqnarray}
\sum_{ \big \{S_j \subseteq \Nbrdis{a}{j} \; : \; j \in \Nbrcl(a)
\backslash \{i\} \big \} } \quad 
\prod_{j \in \Nbrcl(a) \backslash \{i\}} 
\Genmess_{j \ra a}(\unsat{\acl}{j}, S_j) & = & 
\prod_{j \in \Nbrcl(a) \backslash \{i\}} \quad 
\Big[\sum_{ S_j \subseteq \Nbrdis{a}{j} } 
\Genmess_{j \ra a}(\unsat{\acl}{j}, S_j) \Big]
\nonumber \\
\label{EqnPart1}
& = & \prod_{j \in \Nbrcl(a) \backslash \{i\}} \Run{j}{\acl}.
\end{eqnarray}
On the other hand, the weight assigned to the event that each is
either unsatisfying, satisfying or $*$ can be calculated as follows.
Consider a partition $J^u \cup J^s \cup J^*$ of the set $\Nbrcl(a)
\backslash \{i\}$, where $J^u$, $J^s$ and $J^*$ corresponds to the
subsets of unsatisfying, satisfying and $*$ assignments respectively.
The weight $W(J^u, J^s, J^*)$ associated with this partition takes the
form
\begin{equation*}
 \sum_{ \big \{S_j \subseteq \Nbrdis{a}{j} \; : \; j \in J^u \big \} }
\sum_{ \big \{S_j \subseteq \Nbragree{a}{j} \; : \; j \in J^s \big \}
} \quad \prod_{j \in J^u} \Genmess_{j \ra a}(\unsat{\acl}{j}, S_j)
\prod_{j \in J^s} \Genmess_{j \ra a}(\sat{\acl}{j}, S_j) \prod_{j \in
J^*} \Genmess_{j \ra a}(*, \emptyset).
\end{equation*}
Simplifying by distributing the sum and product leads to
\begin{eqnarray*}
W(J^u, J^s, J^*) & = & \prod_{j \in J^u} \Big[ \sum_{ S_j \subseteq
\Nbrdis{a}{j} } \Genmess_{j \ra a}(\unsat{\acl}{j}, S_j) \Big] \quad
\prod_{j \in J^s} \Big[ \sum_{ S_j \subseteq \Nbragree{a}{j} }
\Genmess_{j \ra a}(\sat{\acl}{j}, S_j) \Big] \quad \prod_{j \in J^*}
\Genmess_{j \ra a}(*, \emptyset) \\
& = & \prod_{j \in J^u} \Run{j}{a} \quad \prod_{j \in J^s}
\big[\Rstar{j}{a} - \Genmess_{j \ra a}(*, \emptyset) \big] \quad
\prod_{j \in J^*} \Genmess_{j \ra a}(*, \emptyset),
\end{eqnarray*}
where we have used the definitions of $\Run{j}{a}$ and $\Rstar{j}{a}$
from Section~\ref{AppRform}.  Now summing $W(J^u, J^s, J^*)$ over all
partitions $J^u \cup J^s \cup J^*$ of $\Nbrcl(a) \backslash \{i\}$
yields
\begin{eqnarray}
& \sum \limits_{J^u \cup J^s \cup J^*} & W(J^u, J^s, J^*) \nonumber \\
& = & \sum_{J^u \subseteq \Nbrcl(a) \backslash \{i\}} \; \;  \prod_{j
\in J^u} \Run{j}{a} \quad \sum_{J^s \cup J^* = \Nbrcl(a) \backslash
\{J^u \cup i\} } \Big \{ \prod_{j \in J^s} \big[\Rstar{j}{a} -
\Genmess_{j \ra a}(*, \emptyset) \big] \prod_{j \in J^*} \Genmess_{j
\ra a}(*, \emptyset \Big \} \nonumber \\
& = & \sum_{J^u \subseteq \Nbrcl(a) \backslash \{i\}} \quad \prod_{j \in
J^u} \Run{j}{a} \prod_{j \in \Nbrcl(a) \backslash \{J^u \cup i\}}
\Rstar{j}{a} \nonumber \\
\label{EqnPart2}
& = & \prod_{j \in \Nbrcl(a) \backslash \{i\}} \big[\Run{j}{a} +
\Rstar{j}{a} \big],
\end{eqnarray}
where we have used the binomial identity twice.  Overall,
equations~\eqref{EqnPart1} and~\eqref{EqnPart2} together yield that
\begin{eqnarray*}
\tovarstar{a}{i} & = & \prod_{j \in \Nbrcl(a) \backslash \{i\}}
\big[\Run{j}{a} + \Rstar{j}{a} \big] - \prod_{j \in \Nbrcl(a)
\backslash \{i\}} \Run{j}{\acl},
\end{eqnarray*}
which establishes equation~\eqref{EqnClausetoVarstar}.

\vtiny

(iii) Finally, turning to equation~\eqref{EqnClausetoVarun}, for $x_i
 = \unsat{a}{i}$ and $\Par{i} \subseteq \Nbrdis{a}{i}$, there are only
 two possibilities for the values of $x_{\Nbrcl(a) \backslash \{i\}}$:
\begin{enumerate}
\item[(a)] either there is one satisfying variable and everything else
is unsatisfying, or 
\item[(b)] there are at least two variables that are
satisfying or $*$.
\end{enumerate}
We first calculate the weight $W(A)$ assigned to possibility (a),
again using the BP update equation~\eqref{EqnBPClausetoVar}:
\begin{eqnarray}
W(A) & = & \sum_{k \in \Nbrcl(a)\backslash \{i\}} \sum_{S^k \subseteq
\Nbragree{a}{k}} \Genmess_{k \ra a}(\sat{a}{k}, S^k \cup \{a\})
\prod_{j \in \Nbrcl(a) \backslash \{i, k\} } \sum_{S^j \subseteq
\Nbrdis{a}{j}} \Genmess_{j \ra a}(\unsat{j}{a}, S^j) \nonumber \\
\label{EqnWA}
& = & \sum_{k \in \Nbrcl(a) \backslash \{i\}} \Rsat{k}{a} \prod_{ j
\in \Nbrcl(a) \backslash \{i, k\}} \Run{j}{a},
\end{eqnarray} 
where we have used the definitions of $\Rsat{k}{a}$ and $\Run{k}{a}$
from Section~\ref{AppRform}.

We now calculate the weight $W(B)$ assigned to possibility (b) in the
following way.  From our calculations in part (ii), we found that the
weight assigned to the event that each variable is either
unsatisfying, satisfying or $*$ is $\prod_{j \in \Nbrcl(a) \backslash
\{i\}} \big[\Run{j}{a} + \Rstar{j}{a} \big]$.  The weight $W(B)$ is
given by subtracting from this quantity the weight assigned to the
event that there are \emph{not} at least two $*$ or satisfying
assignments.  This event can be decomposed into the disjoint events
that either all assignments are unsatisfying (with weight $\prod_{j
\in \Nbrcl(a) \backslash \{i\}} \Run{j}{\acl}$ from part (ii)); or
that exactly one variable is $*$ or satisfying.  The weight
corresponding to this second possibility is
\begin{eqnarray*}
\sum_{k \in \Nbrcl(a) \backslash \{i\} } & & \big[\Genmess_{k \ra a}(*,
\emptyset) + \sum_{S^k \subseteq \Nbragree{a}{k}} \Genmess_{k \ra
a}(\sat{k}{a}, S^k) \big] \quad \prod_{j \in \Nbrcl(a) \backslash \{i,
k\} } \sum_{S^j \subseteq \Nbrdis{j}{a}} \Genmess_{j \ra
a}(\unsat{j}{a}, S^j)
\\
& = & \sum_{k \in \Nbrcl(a) \backslash \{i\} } \Rstar{k}{a} \prod_{j
\in \Nbrcl(a) \backslash \{i, k\} } \Run{j}{a}.
\end{eqnarray*}
Combining our calculations so far we have
\begin{eqnarray}
\label{EqnWB}
W(B) & = & \prod_{j \in \Nbrcl(a) \backslash \{i\}} \big[\Run{j}{a} +
\Rstar{j}{a} \big] - \sum_{k \in \Nbrcl(a) \backslash \{i\} }
\Rstar{k}{a} \prod_{j \in \Nbrcl(a) \backslash \{i, k\} } \Run{j}{a} -
\prod_{j \in \Nbrcl(a) \backslash \{i\}} \Run{j}{\acl}.
\end{eqnarray}
Finally, summing together the forms of $W(A)$ and $W(B)$ from
equations~\eqref{EqnWA} and~\eqref{EqnWB} respectively, and then
factoring yields the desired equation~\eqref{EqnClausetoVarun}.
\end{proof}

\section{Proofs for random formulae}

\subsection{Proof of Lemma~\ref{LemCond}}
\label{AppLemCond}

In order to prove~\eqref{eq:cond}, it suffices by the Markov
inequality to show that for every integer $d$ in the interval $[\delta
n, 2 \delta n]$, it holds that 
\begin{equation}
\label{eq:avg_cond}
\frac{\Elexp^n[W^{\phi}[n-n_* = d]]}{\rho^{3n}} = \exp(-\Omega(n)).
\end{equation} 
To establish~\eqref{eq:avg_cond}, consider a fixed set of $d$
variables. The average $W$-weight assigned to the event that this set
of size $d$ constitutes all the non-star variables is bounded by
\[
\rho^{n - d} \sum_{r=0}^d (1-\rho)^{d - r} \binom{d}{r} \binom{\alpha
n}{r} (d/n)^{kr},
\]
where $r$ represents the number of constrained variables.  We obtain
this bound by the following reasoning.  First, the $n-d$ variables
assigned $\ast$ all receive weight $\rho$.  Otherwise, if $r$ out of
the remaining $d$ variables are constrained, there must be $r$ clauses
chosen from a total of $\alpha n$, and each such clause must have all
of its $k$ variables chosen from within the set of $d$ non-star
variables.  

Consequently, the total probability of having $d$ non-star variables
is bounded by
\begin{eqnarray*}
\rho^{n - d} \binom{n}{d} \sum_{r=0}^{d} (1 -\rho)^{d - r}
\binom{d}{r} \binom{\alpha n}{r} \left( \frac{d}{n} \right)^{k r} & \leq &
\rho^{n - d} \left(\frac{e n}{d}\right)^{d} \sum_{r=0}^{d} (1 -
\rho)^{d - r} \left(\frac{ e d}{r}\right)^r \left(\frac{\alpha e
n}{r}\right)^r \left(\frac{d}{n}\right)^{k r} \\
&=& \rho^{n - d} \left(\frac{(1-\rho) e n}{d}\right)^{d}
\sum_{r=0}^{d} \left(\frac{e^2 d^{k+1} \alpha}{r^2 (1-\rho)
n^{k-1}} \right)^r,
\end{eqnarray*}
Recalling that $1-\rho = \delta^2$ and $d \in [\delta n, \, 2 \delta
n]$, we obtain that the last expression is at most
\begin{eqnarray*}
\rho^{n - 2 \delta n} \left( \frac{\delta^2 e n}{\delta n} \right)^{d}
\; \; \sum_{r=0}^{2 \delta n} \left(\frac{e^2 (2 \, \delta n)^{k+1}
\alpha}{r^2 \delta^2 n^{k-1}}\right)^r
& = & \rho^{n - 2 \delta n} (\delta e)^{d} \sum_{r=0}^{2 \delta n}
\left(\frac{e^2 2^{k+1} \delta^{k-1} n^2 \alpha}{r^2}\right)^r \\ &
\leq & \rho^{n-2} (\delta e)^{\delta n} \sum_{r=0}^{2 \delta n}
\left(\frac{2^{k+1} \alpha \delta^{k-1} n^2 e^2}{r^2}\right)^r,
\end{eqnarray*}
where the final inequality is valid when $\delta e < 1$.  A
straightforward calculation yields that the function $g(r) \defn
\left(\frac{2^{k+1} \alpha \delta^{k-1} e^2 n^2}{r^2}\right)^r$ is
maximized at $r^* = \sqrt{2^{k+1} \alpha \delta^{k-1}} n$ and the
associated value is $g(r^*) = e^{2 r^*}$.  Consequently, the sum above
is bounded by
\begin{eqnarray*}
2 \delta n \rho^{n - 2 \delta n} (\delta e)^{\delta n} e^{2 r^*} & = &
2 \delta n \rho^{n - 2 \delta n} \left[\delta \exp\left(1+ \frac{2
r^*}{\delta n} \right)\right]^{\delta n} \\
& = & 2 \delta n \rho^{n - 2 \delta n} \left[\delta \exp\left(1+
\sqrt{2^{k+3} \alpha \delta^{k-3}} \right) \right]^{\delta n} \\
& \leq & 2 \delta n \rho^{n - 2 \delta n} \left[\delta \exp\left(1+
\sqrt{2^{k+3} \alpha}\right)\right]^{\delta n}. \\
\end{eqnarray*}
This expression is exponentially smaller than $\rho^{3n}$ for large
$n$ if
\begin{eqnarray}
\label{EqnFinalInq}
\left[\delta \exp\left(1+ \sqrt{2^{k+3} \alpha}\right)\right]^{\delta}
< \rho^3 = (1 - \delta^2)^3.
\end{eqnarray}
Inequality~\eqref{EqnFinalInq} holds for sufficiently small $\delta >
0$, which establishes the lemma.

\subsection{Proof of Lemma~\ref{lem:sum_product}}
\label{AppLemSumProd}

It will be useful to denote $\prod_{b\in \Nbragree{\acl}{i}}
(1-\messtovar{\bcl}{i} )$ by $P_s(i)$ and $\prod_{b\in
\Nbrdis{\acl}{i}} ( 1 - \messtovar{b}{i} )$ by $P_u(j)$.  With this
notation, the $j$'th term in~\eqref{EqnUpMesstovar} is given by
\begin{eqnarray*}
\frac{\messtoclun{j}{\acl}}{ \messtoclun{j}{\acl} +
\messtoclsat{j}{\acl} + \messtoclstar{j}{\acl}} &=&  
\frac{(1- \rho P_u(j))P_s(j)}{(1- \rho P_u(j))P_s(j) + (1-P_s(j))P_u(j) + P_s(j) P_u(j)} 
\\ &=& 
\frac{(1- \rho P_u(j))P_s(j)}{P_s(j) + P_u(j) - \rho P_s(j) P_u(j)} 
\leq 1-\rho P_u(j).
\end{eqnarray*}
We therefore conclude that 
\[
\messtovar{\acl}{i} \leq \prod_{j \in \Nbrcl(\acl) \backslash \{i\}}
\left(1 - \rho P_u(j)\right).
\]
On the other hand, we have $P_u(j) = \prod_{b\in \Nbrdis{\acl}{i}} ( 1
- \messtovar{b}{i} ) \geq \max\left(0,1 - \sum_{b\in \Nbrdis{\acl}{i}}
\messtovar{b}{i}\right)$, so that
\[ 
1- \rho P_u(j)\leq \min\left(1,
(1-\rho)+\rho \sum_{b\in \Nbrdis{\acl}{i}} \messtovar{b}{i}\right).
\]
This yields the bound $\messtovar{\acl}{i}^{t+1} \leq \prod_{j \in
\Nbrcl(\acl) \backslash \{i\}} \min \left(1, (1-\rho) + \rho
\sum_{b\in \Nbrdis{\acl}{i}} \messtovar{b}{j}^t\right)$, from which
equation~\eqref{eq:sum_product_bound} follows.
\subsection{Proof of Lemma~\ref{lem:bp}}
\label{AppLemBP}
We start by estimating the probability that a vertex is bad by
induction. Let $g_K$ denote the probability that $v$ has more than $K$
children, or that one of $v$'s children has more than $K$ children.
Clearly,
\begin{equation} \label{eq:rec_gk}
g_K \leq (K+1)\Prob[Y \geq K] \leq (K+1)(k-1) \Prob[\Bin(M_c,
\frac{\alpha n}{N_c}) \geq \frac{K}{k-1}] \leq \exp(-\Omega(K)).
\end{equation}
Write $q(m,K) = 1 - p(m,K)$ and note that $q(0,K) = 0$ and $q(1,K)
\leq g_K$. By induction, A vertex can be bad for two reasons: it has
two many descendants in the two levels below it, or it has $2$ bad
descendant in the two levels below it.  We may thus bound the
probability of a vertex being bad as
\begin{equation} \label{eq:branching_process}
q(s,K) \leq g_K + \Prob[\Bin(K^2,q(s-2,K)) \geq 2].
\end{equation}
Note also that  
\begin{equation} \label{eq:rec_bin}
\Prob[\Bin(K^2,q(s-2,K)) \geq 2] \leq K^4 q(s-2,K)^2.
\end{equation}
Combining~\eqref{eq:branching_process} and~\eqref{eq:rec_bin} yields
\begin{equation} \label{eq:final_induction}
q(s,K) \leq g_K + K^4 q(s-2,K)^2.
\end{equation}
By (\ref{eq:rec_gk}) when $K$ is sufficiently large $K^4 (2 g_K)^2 <
g_K$.  Thus when $K$ is sufficiently large, it follows from
equation~\eqref{eq:final_induction} that
\[
q(s,K) \leq 2 g_K
\] 
for all $s$. Finally when $K$ is sufficiently large $p(s,K) \geq 1 - 2
g_K$ for all $s$ and $1 - 2 g_K  \geq 1 - \exp(-\Omega(K))$ as needed.



\bibliography{d_spjournal}

\begin{thebibliography}{10}

\bibitem{AP03}
D.~Achiloptas and Y.~Peres.
\newblock The threshold for random $k$-{SAT} is $2^k 2 \log 2 - o(k)$.
\newblock In {\em Proceedings of FOCS}, pages 223--231, 2003.

\bibitem{AR05}
D.~Achlioptas and F.~Ricci-Tersenghi.
\newblock Clustering in random $k$-sat.
\newblock Manuscript, 2005.

\bibitem{Aurell05}
E.~Aurell, U.~Gordon, and S.~Kirkpatrick.
\newblock Comparing beliefs, surveys, and random walks.
\newblock In {\em Neural Information Processing Systems}, January 2005.

\bibitem{BKMP04}
N.~Berger, C.~Kenyon, E.~Mossel, and Y.~Peres.
\newblock Glauber dynamics on trees and hyperbolic graphs.
\newblock To appear. Extended abstract by Kenyon, Mossel and Peres appeared in
  proceeding of 42nd STOC, 2004.

\bibitem{BMWZ03}
A.~Braunstein, M.~M\'{e}zard, M.~Weigt, and R.~Zecchina.
\newblock Constraint satisfaction by survey propagation.
\newblock Technical report, 2003.
\newblock Preprint at URL:http://lanl.arXiv.org/cond-mat/0212451.

\bibitem{BMZ03}
A.~Braunstein, M.~M\'{e}zard, and R.~Zecchina.
\newblock Survey propagation: an algorithm for satisfiability.
\newblock Technical report, 2003.
\newblock Preprint at URL:http://lanl.arXiv.org/cs.CC/0212002.

\bibitem{BZ04}
A.~Braunstein and R.~Zecchina.
\newblock Survey propagation as local equilibrium equations.
\newblock Technical report, 2004.
\newblock Preprint at URL:http://lanl.arXiv.org/cond-mat/0312483.

\bibitem{CR92}
V.~Chvatal and B.~Reed.
\newblock Mick gets some (the odds are on his side).
\newblock In {\em Proceedings of 33rd FOCS}, Pittsburgh Pennsylvania, 1992.

\bibitem{Cook71}
S.~Cook.
\newblock The complexity of theorem-proving procedures.
\newblock In {\em Proceeding of 3rd STOC}, page 151, 1971.

\bibitem{Cooper90}
G.~Cooper.
\newblock The computational complexity of probabilistic inference using
  {B}ayesian belief networks.
\newblock {\em Artificial Intelligence}, 42:393--405, 1990.

\bibitem{Coughlan02}
J.~Coughlan and S.~Ferreira.
\newblock Finding deformable shapes using loopy belief propagation.
\newblock In {\em European Conference on Computer Vision}, 2002.

\bibitem{Dagum93}
P.~Dagum and M.~Luby.
\newblock Approximate probabilistic reasoning in {B}ayesian belief networks is
  {NP}-hard.
\newblock {\em Artificial Intelligence}, 60:141--153, 1993.

\bibitem{Fer92}
W.~F. de~la Vega.
\newblock On random 2-sat.
\newblock Unpublished manuscript., 1992.

\bibitem{Dechter03}
R.~Dechter.
\newblock {\em Constraint processing}.
\newblock Morgan Kaufmann, Palo Alto, CA, 2003.

\bibitem{DBM00}
O.~Dubois, Y.~Boufkhad, and J.~Mandler.
\newblock Typical random 3-sat formulae and the satisfiability threshold.
\newblock In {\em Proceedings of 11'th SODA}, pages 126--127, 2000.

\bibitem{Feige02}
U.~Feige.
\newblock Relations between average case complexity and approximation
  complexity.
\newblock In {\em Proceedings of the 34th STOC}, 2002.

\bibitem{Freeman00}
W.~T. Freeman, E.~C. Pasztor, and O.~T. Carmichael.
\newblock Learning low-level vision.
\newblock {\em Intl. J. Computer Vision}, 40(1):25--47, 2000.

\bibitem{Friedgut99}
E.~Friedgut.
\newblock Neccesary and sufficient conditions for sharp threhsolds of graph
  properties and the $k$-problem.
\newblock {\em J. Amer. Math. Soc.}, 12:1017--1054, 1999.

\bibitem{Gallager63}
R.~G. Gallager.
\newblock {\em Low-density parity check codes}.
\newblock MIT Press, Cambridge, MA, 1963.

\bibitem{Goe96}
A.~Goerdt.
\newblock A remark on random 2-sat.
\newblock {\em J. Computer System and Sciences}, 53:469--486, 1996.

\bibitem{Janson00}
S.~Janson, T.~{\L}uczak, and A.~Rucinski.
\newblock {\em Random graphs}.
\newblock Wiley-Interscience Series in Discrete Mathematics and Optimization.
  Wiley-Interscience, New York, 2000.

\bibitem{KKL00}
A.~Kaporis, L.~M. Kirousis, and E.~G. Lalas.
\newblock The probabilistic analysis of a greedy satisfiability algorithm.
\newblock In {\em Proceedings of 10'th Annual European Symposium on Algorithm},
  pages 574--585, 2000.

\bibitem{Kirkpatrick04}
S.~Kirkpatrick.
\newblock On survey propagation.
\newblock Personal communication, 2004.

\bibitem{Frank01}
F.~Kschischang, B.~Frey, and H.-A. Loeliger.
\newblock Factor graphs and the sum-product algorithm.
\newblock {\em IEEE Trans. Info. Theory}, 47:498--519, February 2001.

\bibitem{Levin86}
L.~Levin.
\newblock Average case complete problems.
\newblock {\em SIAM Jour. Comput.}, 15, 1986.

\bibitem{MMW05}
E.~Maneva, E.~Mossel, and M.~J. Wainwright.
\newblock A new look at survey propagation and its generalizations.
\newblock In {\em Proceedings of Symposium on Discrete Algorithms}, pages~--,
  2005.

\bibitem{Martinelli99}
F.~Martinelli.
\newblock Lectures on {G}lauber dynamics for discrete spin models.
\newblock In {\em Lectures on probability theory and statistics (Saint-Flour,
  1997)}, volume 1717 of {\em Lecture Notes in Math.}, pages 93--191. Springer,
  Berlin, 1999.

\bibitem{MPZ02}
M.~M\'{e}zard, G.~Parisi, and R.~Zecchina.
\newblock Analytic and algorithmic solution of random satisfiability problems.
\newblock {\em Science}, 297, 812, 2002.
\newblock (Scienceexpress published on-line 27-June-2002;
  10.1126/science.1073287).

\bibitem{MRZ03}
M.~M\'{e}zard, F.~Ricci-Tersenghi, and R.~Zecchina.
\newblock Two solutions to diluted $p$-spin models and xorsat problems.
\newblock {\em J. Stat. Phys.}, 111:505, 2003.

\bibitem{MZ02}
M.~M\'{e}zard and R.~Zecchina.
\newblock Random k-satisfiability: from an analytic solution to an efficient
  algorithm.
\newblock {\em Phys. Rev. E}, 66, 2002.

\bibitem{MZ97}
R.~Monasson and R.~Zecchina.
\newblock Statistical mechanics of the random $k$-satisfiability model.
\newblock {\em Phys. Rev. E}, 3:1357--1370, 1997.

\bibitem{MMZ05}
T.~Mora, M.~M\'{e}zard, and R.~Zecchina.
\newblock Clustering of solutions in the random satisfiability problem.
\newblock {\em Phys. Rev. Lett.}, 2005.
\newblock In press.

\bibitem{Parisi02}
G.~Parisi.
\newblock On local equilibrium equations for clustering states.
\newblock Technical report, 2002.
\newblock Preprint at URL:http://lanl.arXiv.org/cs.CC/0212047.

\bibitem{Pearl88}
J.~Pearl.
\newblock {\em Probabilistic reasoning in intelligent systems: networks of
  plausible inference}.
\newblock Morgan Kaufmann, Palo Alto, CA, 1988.

\bibitem{Richardson01a}
T.~Richardson and R.~Urbanke.
\newblock The capacity of low-density parity check codes under message-passing
  decoding.
\newblock {\em IEEE Trans. Info. Theory}, 47:599--618, February 2001.

\bibitem{RPF99}
J.~Rosenthal, J.~Plotkin, and J.~Franco.
\newblock The probability of pure literals.
\newblock {\em Journal of Computational Logic}, 9:501--513, 1999.

\bibitem{SKC93}
B.~Selman, H.~Kautz, and B.~Cohen.
\newblock Local search strategies for satisfiability testing.
\newblock In D.~S. Johnson and M.~A. Trick, editors, {\em Cliques, coloring,
  and satisfiability : second DIMACS implementation challenge, October 11-13,
  1993}, Providence, RI, 1996. American Mathematical Society.

\bibitem{Stanley1}
R.~P. Stanley.
\newblock {\em Enumerative combinatorics}, volume~1.
\newblock Cambridge University Press, Cambridge, UK, 1997.

\bibitem{TJ02}
S.~Tatikonda and M.~I. Jordan.
\newblock Loopy belief propagation and {G}ibbs measures.
\newblock In {\em Uncertainty in Artificial Intelligence (UAI), Proceedings of
  the Eighteenth Conference}, 2002.

\bibitem{WJ03}
M.~J. Wainwright and M.~I. Jordan.
\newblock Graphical models, exponential families, and variational methods.
\newblock Technical report, UC Berkeley, Department of Statitics, No. 649,
  2003.
\newblock Preprint at
  URL:http://www.eecs.berkeley.edu/$\widetilde{\phantom{e}}$wainwrig/Papers/Wa%
iJorVariational03.ps.

\bibitem{WM05}
M.~J. Wainwright and E.~Maneva.
\newblock Lossy source encoding via message-passing and decimation over
  generalized codewords of {LDGM} codes.
\newblock In {\em Proceedings of IEEE International Symposium on Information
  Theory}, 2005.

\bibitem{Wang97}
J.~Wang.
\newblock Average case computational complexity theory.
\newblock In L.~Hemaspaandra and A.~Selman, editors, {\em Complexity theory
  retrospective}, volume~II. Springer, 1997.

\bibitem{Welling01}
M.~Welling and Y.~Teh.
\newblock Belief optimization: A stable alternative to loopy belief
  propagation.
\newblock In {\em Uncertainty in Artificial Intelligence}, July 2001.

\bibitem{Yedidia05}
J.~Yedidia, W.~T. Freeman, and Y.~Weiss.
\newblock Constructing free energy approximations and generalized belief
  propagation algorithms.
\newblock {\em IEEE Trans. Info. Theory}, 51(7):2282--2312, July 2005.

\bibitem{YFW03}
J.~S. Yedidia, W.~T. Freeman, and Y.~Weiss.
\newblock Understanding belief propagation and its generalizations.
\newblock In {\em Exploring Artificial Intelligence in the New Millennium},
  chapter~8. Science and Technology books, 2003.

\bibitem{Yuille01}
A.~Yuille.
\newblock {CCCP} algorithms to minimize the {B}ethe and {K}ikuchi free
  energies: {C}onvergent alternatives to belief propagation.
\newblock {\em Neural Computation}, 14:1691--1722, 2002.

\end{thebibliography}


\end{document}